\documentclass[aps,pra,superscriptaddress, twocolumn, preprintnumbers,amsmath,amssymb,nobalancelastpage,10pt,longbibliography]{revtex4-1}

\usepackage{amsmath}
\usepackage{verbatim}
\usepackage{graphicx}
\usepackage{color}
\usepackage{braket}
\usepackage{ bbold }
\usepackage{ bbold }
\usepackage{yfonts}
\usepackage[normalem]{ulem}

\usepackage{graphicx}% Include figure files
\usepackage{dcolumn}% Align table columns on decimal point
\usepackage{bm}% bold math
\usepackage[dvipsnames]{xcolor}

\usepackage[colorlinks=true,citecolor=blue,linkcolor=blue,urlcolor=blue]{hyperref}
\usepackage{subfigure}
\definecolor{darkgreen}{rgb}{0,0.60,.2}

\begin{document}

\title{Topological properties of the long-range Kitaev chain with Aubry-Andr\'e-Harper modulation}

\author{Joana Fraxanet}
 \affiliation{ICFO - Institut de Ci\`encies Fot\`oniques, The Barcelona Institute of Science and Technology, 08860 Castelldefels (Barcelona), Spain}
\author{Utso Bhattacharya}
 \affiliation{ICFO - Institut de Ci\`encies Fot\`oniques, The Barcelona Institute of Science and Technology, 08860 Castelldefels (Barcelona), Spain}
\author{Tobias Grass}
 \affiliation{ICFO - Institut de Ci\`encies Fot\`oniques, The Barcelona Institute of Science and Technology, 08860 Castelldefels (Barcelona), Spain}
\author{Debraj Rakshit}
 \affiliation{ICFO - Institut de Ci\`encies Fot\`oniques, The Barcelona Institute of Science and Technology, 08860 Castelldefels (Barcelona), Spain}
\author{Maciej Lewenstein}
 \affiliation{ICFO - Institut de Ci\`encies Fot\`oniques, The Barcelona Institute of Science and Technology, 08860 Castelldefels (Barcelona), Spain}
\affiliation{ICREA, Pg. Lluis Companys 23, 08010 Barcelona, Spain}
\author{Alexandre Dauphin}
 \affiliation{ICFO - Institut de Ci\`encies Fot\`oniques, The Barcelona Institute of Science and Technology, 08860 Castelldefels (Barcelona), Spain}

\date{\today}

\begin{abstract}
We present a detailed study of the topological properties of the Kitaev chain with long-range pairing terms and in the presence of an Aubry-Andr\'e-Harper on-site potential. Specifically, we consider algebraically decaying superconducting pairing amplitudes; the exponent of this decay is found to determine a critical pairing strength, below which the chain remains topologically trivial. Above the critical pairing, topological edge modes are observed in the central gap. For sufficiently fast decay of the pairing, these modes are identified as Majorana zero-modes. However, if the pairing term decays slowly, the modes become massive Dirac modes.
Interestingly, these massive modes still exhibit a true level crossing at zero energy, which points towards an initimate relation to Majorana physics.
We also observe  a clear lack of bulk-boundary correspondence in the long-range system, where bulk topological invariants remain constant, while dramatic changes appear in the behavior at the edge of the system. 
In addition to the central gap around zero energy, the Aubry-Andr\'e-Harper potential also leads to other energy gaps at non-zero energy. 
As for the analogous short-range model, the edge modes in these gaps can be characterized through a 2D Chern invariant. However, in contrast to the short-range model, this topological invariant does not correspond to the number of edge mode crossings anymore. This provides another example for the weakening of the bulk-boundary correspondence occurring in this model. Finally, we discuss possible realizations of the model with ultracold atoms and condensed matter systems.

\end{abstract}

\maketitle

\section{\label{sec:level1} Introduction}

Topological phases of matter have now been established as an intriguing research area in condensed matter physics~\cite{hasan10, qi11, bernevig13,asboth16,dalibard19}. Non-interacting symmetry protected topological (SPT) phases are either band insulators or superconductors: they have a gap in their energy spectrum of the bulk and are characterized by a global order parameter, called topological invariant. These topological invariants are integers. As a consequence, they cannot be smoothly deformed, within the same symmetry class of the Hamiltonian, to another phase labelled by a different value of the invariant. Therefore, topological phase transitions are always accompanied by the closing and subsequent reopening of the bulk gap.  A classification of SPT phases in terms of their symmetries (time-reversal, particle-hole and chiral) and their dimensionality has been performed in the celebrated "periodic table of topological insulators and superconductors"~\cite{altland97,schnyder08,schnyder09,kitaev09,chiu16}. Each class is characterized by a different topological invariant which can take integer $\mathbb{Z}$, even-integer $2\mathbb{Z}$  or binary $\mathbb{Z}_2$ values. Finally, SPT phases possess gapless boundary states protected against local perturbations and the number of these protected surface states is proportional to the topological invariant.

The superconducting Kitaev chain~\cite{kitaev01} is a prime example of a 1D non-interacting topological system. The main constituents of the $p$-wave Kitaev chain~\cite{kitaev01} are spinless fermions that undergo short-range hoppings, primarily between nearest neighbors, in a 1D lattice.  They can also get paired together into superconducting Cooper pairs with opposite momenta. The Kitaev chain exhibits time-reversal, chiral and particle-hole symmetry; thus it belongs to the BDI class of the topological classification~\cite{altland97}. This model can host non-local Majorana modes (MMs), i.e. zero-energy modes localized at the two boundaries of the chain. Such system is also characterized by a topological invariant, called winding number, and the number of MMs at each end of the chain is equal to this winding number. Therefore, these MMs are topologically protected against local perturbations and, hence, cannot be removed without a topological phase transition, i.e. a transition from a topological superconducting phase to the trivial superconducting phase when all MMs become paired.

The robustness of the MMs makes them promising  candidates to build topological quantum computers. They could be used as qubits that can store and manipulate quantum information in a decoherence free manner \cite{kitaev01}. Recently, there have been many proposals for the realisation of  MMs in different systems like heterostructures of topological insulators and $s$-wave superconductors~\cite{liang08} or cold fermion systems near $p$-wave Feshbach resonance \cite{tewari07}, or with attractive $s$-wave interaction in the presence of Rashba spin-orbit coupling and Zeeman field~\cite{zhang08,sato09}. Other proposals include heterostructures of spin-orbit-coupled semiconductor thin films~\cite{sau10a, sau10b}, or nanowires~\cite{sau10b, lutchyn10, oreg10}  coupled via proximity effect with s-wave superconductors and a Zeeman field.  Majorana fermions may also emerge from magnetic nanoparticles on a superconductor without spin-orbit coupling~\cite{choy11,gorczyca19}.

The short-range Kitaev chain is a version of the integrable $p$-wave superconducting chain of fermions.  The primary focus of the present work is a long-range Kitaev chain, in which the superconducting pairing term decays with distance as a power-law~\cite{vodola14, vodola15, viyuela16, lepori17, bhattacharya19}. While the topological phases of the short-range Kitaev chain are characterized by $\mathbb{Z}$ topological invariants, the present  model  posseses a new unconventional topological phase characterized by half-integer $\mathbb{Z}/2$ winding numbers, even though the system belongs to the same BDI class and respects the same discrete symmetries of the short-range model. For slow power law decay, i.e. for decay exponent $\alpha$ given by $\alpha\ll 1$, the MMs of the short-range model ($\alpha\gg 1$) coalesce to form massive nonlocal edge states called massive Dirac modes (MDM).  These new edge states lie within the bulk gap and are topologically robust against local perturbations that do not violate fermionic parity and particle-hole symmetry.  Thus, like the MMs, MDMs  may also find novel applications in the area of topological quantum computations. 

One must note that the quantum critical behavior of systems with long-range interactions/pairings  can be very different from that of the short-range ones, due to the non-local nature of the interactions. Therefore, over the recent years, much effort has been devoted to understanding the static~\cite{koffel12, grass14} and dynamic properties~\cite{nandy18} of  long-range systems through quantum quenches~\cite{hauke13,dutta17,lorenzo18}, periodic drives~\cite{bhattacharya19}, through establishment of the Lieb-Robinson bounds~\cite{fossfeig15}, and through studies of the growth of entanglement entropy~\cite{buyskikh16}, Bell inequalities~\cite{piga19}, and thermalization~\cite{van16,bhattacharya18} (for a detailed review, see~\cite{maity19}).

Interestingly, although there have been numerous recent claims regarding the detection of Majorana fermions in superconducting/semiconducting heterostructures~\cite{mourik12, rokhinson12, deng12, das12, churchill13, finck13, stanescu13, jia17}, the quest for Majorana fermions has been quite challenging and even controversial. A major issue that needs to be addressed in any system hosting MMs or MDMs is the presence of correlated or uncorrelated local inhomogeneities in the onsite potential, which may alter the physics of such systems. 

In recent years, new experimental techniques in the field of ultracold atoms have emerged, where a superposition of two periodic lattices with commensurate or incommensurate periods can be created~\cite{luschen18, modugno10, roati08, lohse16, nakajima16, an20}. In particular, this allows for the experimental implementation of theoretical models such as the Aubry-Andr\'e-Harper (AAH) model~\cite{aubry80,harper55} and opens the road for the study of the role of spatial inhomogeneities in close connection to the experiments discussed above. The AAH model raised great interest as, for incommensurate lattices, the system is quasiperiodic  and exhibits a localization-delocalization transition due to its self-dual nature~\cite{anderson58,aubry80,soukoulis82,sarma88}. More recently, the AAH model and similar quasiperiodic models have also been extensively studied for their topological properties~\cite{kraus2012,kraus2012a,tanese2014,dareau2017,madsen13}. In fact, the AAH Hamiltonians can be seen as the dimensional reduction of a 2D Hofstadter model~\cite{hofstadter76}, which describes electrons in a 2D lattice subjected to a perpendicular magnetic field akin to that of a quantum Hall system. Both models, the AAH chain and the 2D Hofstadter lattice, are described by a set of Harper equations~\cite{harper55}, and the two momentum variables of the 2D model map onto one momentum variable and a $2\pi$-peridodic Hamiltonian parameter in the 1D model. Accordingly, the AAH model also exhibits a fractal Hofstadter butterfly energy spectrum, and can host topologically protected boundary modes reminiscent of the 2D system;  it is in turn characterized by a 2D topological invariant called Chern number.

The interplay between nearest neighbor $p$-wave superconducting pairing and the AAH potential has been studied in Refs.~\cite{zeng19, zeng18, yahyavi19}, for both the commensurate and the incommensurate cases. One of the key observations in such systems has been that a finite amount of superconductivity is required for unpaired Majorana modes to be present in the system~\cite{motrunich01}. Although comprehensive studies~\cite{degottardi13a,degottardi13b, yahyavi19} have been performed on unifying the topological phase diagram for a range of periodic, incommensurate and disordered potentials in the short-range Kitaev chain, the combined effect of long-range pairings and AAH onsite potential on the topological phase diagram of the Kitaev chain has never been explored. Here, we present a comprehensive study of the 1D long-range Kitaev Hamiltonian with AAH type spatially inhomogeneous potential. Through extensive numerical calculations of the topological invariants and edge spectra, we reveal how the critical behavior changes throughout the topological phase diagram. Specifically, we show that, for sufficiently slowly decaying pairing, the central bands are characterized by the same half-integer topological invariant as in the long-range Kitaev chain with homogeneous chemical potential. The edge modes appear at finite energy, which suggests to interpret them as massive Dirac modes (MDMs). However, the presence of the AAH potential induces a non-monotonous dispersion of these modes when tuning the chemical potential or the superconducting pairing strength. This leads to a crossing of these levels in isolated points at zero energy, suggesting an intrinsic relation to Majorana physics. At the same time, none of this interesting behavior at the system edge is reflected by the topological invariant, suggesting that the system exhibits a weakened bulk-boundary correspondence. This conclusion is corroborated by studying the behavior in higher band gaps induced by the AAH potential. Specifically, we find that  long-range pairing removes edge state crossings by turning them into avoided crossings, whereas the topological invariants of the corresponding bands remain unchanged.

\noindent{\bf Plan of the paper.} Section~\ref{sec:level2} introduces the model and discusses its general properties. Section~\ref{sec:level3} is devoted to the discussion of the winding number and challenges related to the numerical computation of such winding number in the presence of long-range pairing. In Section~\ref{sec:level4}, we analyze the effects of the AAH modulation on the central gap behavior of the long-range Kitaev chain. This includes characterizing the MMs and MDMs and their corresponding winding numbers, studying the existence of a critical superconducting pairing, and a detailed analysis of the level crossings at zero energy. Section~\ref{sec:level5} is devoted to the effect of the long-range pairings on higher energy gaps, which contain edge states arising from the AAH potential. Finally, in Section~\ref{sec:level6} we discuss possible experimental realizations of the model in condensed matter and in systems of ultracold atoms or molecules.

\section{\label{sec:level2} Model Hamiltonian}

We consider a chain of spinless fermions with an onsite AAH modulation of the chemical potential and a long-range $p$-wave superconducting pairing. Its Hamiltonian can be written as
\begin{eqnarray}\label{eq: hamiltonian}
    H =\sum_{i=0}^{N-1} \biggl[&&-t \left (c^\dagger_{i+1}c_{i} + c^\dagger_{i}c_{i+1}\right) - \mu f(i)\left ( 2c^\dagger_{i}c_{i}-1\right)  \nonumber\\
    &&+\sum_{l=1}^{N-1}\frac{\Delta}{l^\alpha}\left (c^\dagger_{i+l}c^\dagger_{i} + c_{i}c_{i+l}\right)\biggl],
\end{eqnarray}
where $t$ is the hopping amplitude, $\mu$ is the chemical potential, $\Delta$ is the superconducting pairing amplitude and $c_{i}(c^\dagger_{i})$ are the annihilation (creation) operators at the $i$-th site of the chain. The superconducting pairing is taken to decay as a power law $l^{-\alpha}$, where $l$ denotes the distance between the sites and the scaling exponent $\alpha \in \mathbb{R}$~\footnote{we here set the lattice spacing to unity}. For a constant chemical potential $f(i)=1$, we recover the long-range Kitaev chain with homogeneous onsite potential. As mentioned above, this model is known to be topologically equivalent to the short-range Kitaev chain with nearest-neighbor pairing terms for $\alpha \gg 1$. In contrast, for $\alpha \ll 1$ the model can host a long-range topological phase with MDMs, characterized by a half integer topological invariant. 

We also consider an AAH onsite chemical potential
\begin{equation}\label{eq:modulation}
f(i) = \cos (2 \pi \beta i + \phi ),
\end{equation}
 where the modulation frequency $\beta$ gives the periodicity of the potential and the parameter $\phi$ shifts the origin of the modulation. In particular, for irrational $\beta$, the system becomes incommensurate with respect to the lattice periodicity. The latter can lead to interesting effects such as gap openings or localization-delocalization transitions~\cite{aubry80,soukoulis82,sarma88}. In this paper, we consider a system with commensurate values of $\beta= p/q$, and characterize its topology. Both the AAH model and the Kitaev model are interesting due to their topological properties separately from each other. Here, we study the interplay between the two.
 
The AAH potential has periodicity of $q$ sites and we therefore consider a system of $N$ sites with $L$ supercells, where $N=qL$. First, we write the Hamiltonian in the Bogoliubov-de-Gennes (BdG) basis to properly treat the pairing term. The Hamiltonian of Eq.~\eqref{eq: hamiltonian} then reads
\begin{eqnarray}\label{eq: hamiltonian rs}
    H = \sum_{u=0}^{L-1} \biggl[&& \chi_u^\dagger H_{\text{local}} \chi_u + \left(\chi_u^\dagger H_{\text{hop}} \chi_{u+1} + \text{h.c.}\right) \nonumber \\
    && + \sum_{l=1}^{L-1}\left(\chi_u^\dagger H_{l} \chi_{u+l} + \text{h.c.}\right)\biggl],
\end{eqnarray}
where $\chi_u = \left(c_{qu}, c^\dagger_{qu}, ..., c_{qu+(q-1)}, c^\dagger_{qu+(q-1)}\right)^T$ is the basis in real space within a supercell denoted by $u$.
$H_\text{local}$ stands for the elements of the Hamiltonian involving operators within the supercell. $H_\text{hop}$ contains the hopping terms between supercells. Finally, $H_l$ includes the long-range pairing terms. The explicit form of each contribution can be found in Appendix~\ref{appendix:construction}. Moreover, we can either impose open boundary conditions (OBC) to study the edge states, or anti-periodic boundary conditions (APBC) assuming $\chi_{u+L} = -\chi_{u}$ to study the bulk properties of the system (see Appendix~\ref{appendix:APBC} for details).

To focus on the bulk properties of the system, let us impose APBC and compute the Hamiltonian in the Fourier space, which reads
\begin{eqnarray}
    H =&& \sum_{k} \biggl[ \chi_k^\dagger H_\text{local} \chi_k + \left(e^{ik}\chi_k^\dagger H_\text{hop} \chi_k + \text{h.c.}\right) \nonumber \\ && + \sum_{l=1}^{L-1}\left(e^{ikl}\chi_k^\dagger H_{l} \chi_{k} + \text{h.c.}\right)\biggl],
\end{eqnarray}
where $\chi_k$ denotes the Fourier transform of the supercell vector, with the Fourier sum being performed over the quasimomenta $k = (2m + 1)\pi/L$, $ 0 \leq m \leq L-1$.
Finally, for the thermodynamic limit $(L \rightarrow \infty)$, the Hamiltonian reads as
\begin{eqnarray}\label{eq: hamiltonian inf}
    H = \sum_{k} \biggl[ &&\chi_k^\dagger H_\text{local} \chi_k + \left(e^{ik}\chi_k^\dagger H_\text{hop} \chi_k + \text{h.c.}\right) \nonumber \\ && + \left(\chi_k^\dagger H_\text{inf} \chi_{k} + \text{h.c.}\right)\biggl].
\end{eqnarray}
In order to better understand the contribution of the long-range pairing, it is worth looking into the definition of $H_\text{inf}$, which takes the form (see App.~\ref{appendix:construction} for more details on the derivation)
\begin{equation}
H_{\text{inf}} = 
\begin{pmatrix}
C_{0,0} & C_{0,1}  & \cdots & C_{0,q-1}\\
C_{1,0} & C_{1,1} &  \cdots  & C_{1,q-1}\\
\vdots   & \vdots  & \ddots & \vdots\\
C_{q-1,0} & C_{q-1,1} &  \cdots & C_{q-1,q-1}
\end{pmatrix},
\end{equation}
where $C_{x,y} = ig_{xy}\sigma_y$. The function $g_{xy}$ is defined as follows
\begin{eqnarray}\label{eq:HLP}
g_{xy} &&= -\sum_{l=1}^{\infty} \frac{\Delta e^{ikl}}{[lq-(x-y)]^\alpha}  = \frac{e^{ik}}{q^{\alpha}} \text{HLP}_{\alpha} \left(k, q, x-y\right),
\end{eqnarray}
where HLP stands for the Hurwitz-Lerch-Phi function, also called the Lerch transcendent function~\cite{kanemitsu00, vodola14}:

\begin{equation}
\text{HLP}_{\alpha}\left(k, q, x-y\right) = \sum_{l=0}^{\infty}  \frac{e^{ikl}}{[l + \frac{q + (x-y)}{q}]^\alpha}.
\end{equation}

 The HLP function has a singularity at $k=0$ for $\alpha<1$. This singularity encodes the long-range character of the Hamiltonian and contributes both to the edge dynamics and to the bulk topology of the system, creating a long-range unconventional topological phase. Our goal will be to characterize it.

\section{\label{sec:level3} Winding number}

As mentioned in Section \ref{sec:level1}, the model under study lies in the BDI symmetry class of topological insulators and superconductors, that is, it is particle-hole, time-reversal and chiral symmetric. Therefore, in order to characterize the bulk topology of the model, we compute the winding number considering three methods. The first one is the Fukui-Hatsugai-Suzuki algorithm~\cite{fukui05}, which, as we will see, fails to characterize the unconventional long-range topological phase. Secondly, we consider the real-space winding number for finite systems, which converges to the expected value of the winding number for large systems. Finally, we implement an algorithm which relies on the chiral symmetry of the Hamiltonian. For this approach, we make use of the analytical expression for the infinite system from Eq.~\eqref{eq: hamiltonian inf}. 

Before getting into the details of each method, we would like to stop and briefly analyze the long-range case $\alpha \ll 1$ \cite{vodola14, vodola15, viyuela16, lepori17, bhattacharya19}. On the one hand, the long-range character of the Hamiltonian affects the edge states of the system: the MMs hybridize and become massive and form MDMs. On the other hand, the bulk topology of the model hosts two different unconventional topological phases, characterized by a half-integer $\mathbb{Z}/2$ winding numbers. It is important to note that the winding number for short-range systems can only take integer $\mathbb{Z}$ values~\cite{altland97,schnyder08,schnyder09,kitaev09,chiu16}. Nevertheless, the long-range character of the Hamiltonian allows for a new type of characterization of the bulk topology. We will see that the bulk topology of our system is well-defined, but not directly connected to the edge dynamics. In fact, the bulk-edge correspondence gets weakened for long-range systems~\cite{lepori17}.

\subsection{\label{sub:level31} Fukui-Hatsugai-Suzuki algorithm}

A common approach to compute the winding number is the algorithm introduced by Fukui, Hatsugai and Suzuki~\cite{fukui05}. The main idea is to discretize the Brillouin zone of the system $\{k_1, ..., k_s\}$, and then to construct the tensor $U^{(i)}_{m,n} = \langle \psi_m(k_{i+1})\vert\psi_n (k_i) \rangle$ for every $k_i$. Here $\psi_{m(n)}$ is the $m (n)$-th eigenvector of the Hamiltonian. The winding number $\nu$ is then defined as
\begin{equation}\label{eq: Fukui}
 \nu = \frac{1}{\pi} \text{Im} \left[ \log \left(\prod_{i=1}^s  \frac{\vert U^{(i)}\vert}{\sqrt{\vert U^{(i)} \vert \vert U^{(i)} \vert^*}}\right)\right],
\end{equation}
where $\vert \cdot \vert$ denotes the determinant of the tensor $U$. To compute this quantity, we work with periodic boundary conditions. This means that we need to close the loop by computing the scalar product between $\vert\psi_m(k_s)\rangle$ and $\vert\psi_n(k_1)\rangle$.

Recall that the long-range character of the Hamiltonian is encoded in the HLP function in Eq.~\eqref{eq:HLP}, and this function is singular at $k=0$ for $\alpha<1$. That is why, for the case of $\alpha < 1$, one needs to consider a discretization of the Brillouin zone without an explicit inclusion of $k=0$ in order to avoid the singularity. Nevertheless, even if one considers an appropriate discretization of the Brillouin zone, this algorithm always fails to characterize the long-range topological phases through half-integer $\mathbb{Z}/2$ winding numbers. In fact, it can be proven that this algorithm always leads to an integer $\mathbb{Z}$ winding number. In order to understand why, let us consider the case of $f(i) = \text{cst}$. This case is  straightforward as the model has only two bands. However, this argument can be generalized to the multi-band (non-Abelian) case, as required, for investigating the properties that we are interested in [see Eq.~\eqref{eq: Fukui}].
The Zak phase of a single band $n$ is given by the sum of the relative phases over the Brillouin zone
\begin{equation}
c_n = \frac{1}{2\pi i} \sum_{l=1}^{s} F_{1}(k_l),
\end{equation}
where
\begin{equation}
F_{1}(k_l) =  \ln \frac{\langle \psi_n(k_{l+1})\vert\psi_n (k_l) \rangle}{\vert\langle \psi_n(k_{l+1})\vert\psi_n (k_l) \rangle\vert}.
\end{equation}
Note that the function $F_{1}(k_l)$ is periodic along the first Brillouin zone, meaning that $F_{1}(k_l + s\delta_k) = F_{1}(k_l)$, where $\delta_k$ is the spacing of the discretization and $s$ are the number of points considered. Moreover, $F_1$ takes values within the principal branch of the logarithm, such that $-\pi < F_{1}(k_l)/i < \pi$. This means that $F_1(k_l)$ can be written in terms of forward differences and an integer-valued field $n_1$
\begin{equation}
F_1(k_l) = f(k_l + \delta_k)-f(k_l) + 2\pi i n_1(k_l).
\end{equation}
If one closes the loop around the Brillouin zone, the finite differences cancel out, and the Chern number is always integer valued
\begin{equation}
c_n = \sum_{l=1}^s n_1(k_l).
\end{equation}
The only way to obtain non-integer winding numbers would be either to neglect the contribution of $F_{1}(k_s)$, therefore not closing the loop, or to include the point $k=0$. Neither option is reliable, since the first one results in gauge dependent quantities and the second one relies on a singularity point.

\subsection{\label{sub:level32} Real-space winding number}

A second approach to compute the winding number is based on the real-space Hamiltonian of a finite system with $N=qL$ sites~\cite{mondragon14}. In this case, the winding number is defined as follows
\begin{equation}
\nu = \text{Tr}\left(Q_{BA}[X, Q_{AB}]\right),
\end{equation}
where $\text{Tr}$ is the trace per unit cell and $Q_{AB}=\Gamma_A Q \Gamma_B$. The quantity $Q$ is defined as $Q = P_{+}-P_{-}$, where $P_{+}$ and $P_{-}$ are the projectors on the positive or negative eigenstates respectively $P_{\pm} = \sum_{\pm} \vert \psi_{\pm}\rangle\langle \psi_{\pm}\vert$. $\Gamma_A$ and $\Gamma_B$ are the projector operators onto the $A$ and $B$ sublattices of the system~\cite{maffei18}, namely $\Gamma_A = \text{diag}(1,0,1,0\cdots)$ and $\Gamma_B = \text{diag}(0,1,0,1\cdots)$ and $X$ is the position operator.

We emphasize that computing $[X, Q_{AB}]$ is not straightforward for a system with periodic or anti-periodic boundary conditions as the position operator is not periodic. We here use the approach introduced in Ref.~\cite{song14}, for which we consider a system with a domain $ -N/2 \leq x_i \leq N/2$, where $x_i$'s are the eigenvalues of the position operator $X$. Then, the commutator can be written as
\begin{equation}\label{eq:approx_commutator}
     [X, Q_{AB}] = \sum_{\lambda \neq 1} c_{\lambda} \lambda^X Q \lambda^{-X},
\end{equation}
where $\lambda$ are the solutions to $z^{2N+1} = 1$ and $c_{\lambda} = \frac{\lambda^{N+1}}{1-\lambda}$. For this approximation to be effective, the system has to be sufficiently large.

\subsection{\label{sub:level33} Infinite system winding number}

Finally, a third approach to compute the winding number relies on the definition of the winding number for the infinite system. Since the system is chiral, one can always find a basis in which the Hamiltonian is block-off-diagonal (see Appendix~\ref{appendix:block-off})

\begin{equation}\label{eq:block-off}
    H_\text{inf} = \begin{pmatrix} 0 & h \\ h^\dagger &0\end{pmatrix}.
\end{equation}
For such Hamiltonian, the chiral operator is diagonal
\begin{equation}
    \Gamma = \begin{pmatrix} \mathbb{1} & 0 \\ 0 & -\mathbb{1} \end{pmatrix},
\end{equation}
and the winding number can be computed with the help of the expression~\cite{maffei18}
\begin{equation}\label{eq: wn chiral int}
    \nu = \int_0^{2\pi} \frac{1}{2\pi i} \text{Tr}\left[h^{-1}\partial_k h\right], 
\end{equation}
where $h$ is the upper-right block of the Hamiltonian in Eq.~\eqref{eq:block-off}. In our case, we are interested in using the discrete version of Eq.~\eqref{eq: wn chiral int}, which takes the form
\begin{equation}\label{eq:wn chiral}
    \nu = \sum_k \frac{1}{L} \text{Tr}\left[h^{-1}\partial_k h\right].
\end{equation}
Finally, note that $h$ contains the HLP function defined in Eq.~\eqref{eq:HLP}, which is singular at k = 0. In Eq.~\eqref{eq:wn chiral}, the winding number can still be computed as long as one considers a discretization of the Brillouin zone without an explicit inclusion of $k=0$. It is important to note that the discontinuity does not affect the calculations in the continuous limit and therefore the integral in Eq.~\eqref{eq: wn chiral int} is well defined. Note that the same happens to the long-range Kitaev chain with no AAH modulation \cite{bhattacharya19}.
In order to compute the winding number, we also need the analytical derivative of this function

\begin{eqnarray}
\partial_k \text{HLP}_{\alpha}\left(k, q,x-y\right) &&= i e^{\frac{ik}{q^{\alpha+1}}} \big[q \text{HLP}_{\alpha-1}\left(k, q, x-y\right) \nonumber \\
&& +(x-y)\text{HLP}_{\alpha}\left(k, q, x-y\right)\big].
\end{eqnarray}

We will use the winding number computation explained in this section to study the central gap of the model in the following section.

\section{\label{sec:level4}  Central gap of the Kitaev chain with AAH potential: a quantitative study}

The AAH potential splits the energy spectrum of the Kitaev chain into several bands. In this Section, we focus on the behavior of such system around its central gap, that is, around $E=0$. Specifically, we compare the behavior of short-ranged and long-ranged Kitaev chains in the presence or absence of the AAH modulation.
 
In this context, we first study the energy spectra as a function of the chemical potential $\mu$, which reveal the existence of MMs and MDMs for different values of the decay exponent $\alpha$. Then, we characterize these different phases using the real-space \ref{sub:level32} and the infinite winding numbers \ref{sub:level33}. Here, we also compare the accuracy of the two methods. We then investigate how energy spectra and winding numbers depend on the superconducting pairing term $\Delta$, and find that the presence of an AAH modulation leads to a critical superconducting pairing $\Delta_C$: below this critical value, we observe neither MMs nor MDMs in the system. The value of $\Delta_C$ depends on the decay exponent $\alpha$ as well as the amplitude of the chemical potential $\mu$. Finally, we discuss a peculiar behavior found in the presence of both long-range pairing and AAH potential: while the long-range nature of the pairing leads to a degeneracy splitting of the MMs and turns them into MDMs, the AAH is responsible for a true level crossing of these modes which allows them to re-combine at isolated points, and which yields an adiabatic connection between MM phase and MDM phases. 

\subsection{\label{sub:level41}Effect of the AAH potential on the energy spectrum}

\begin{figure}[t]
\includegraphics[width=1\columnwidth]{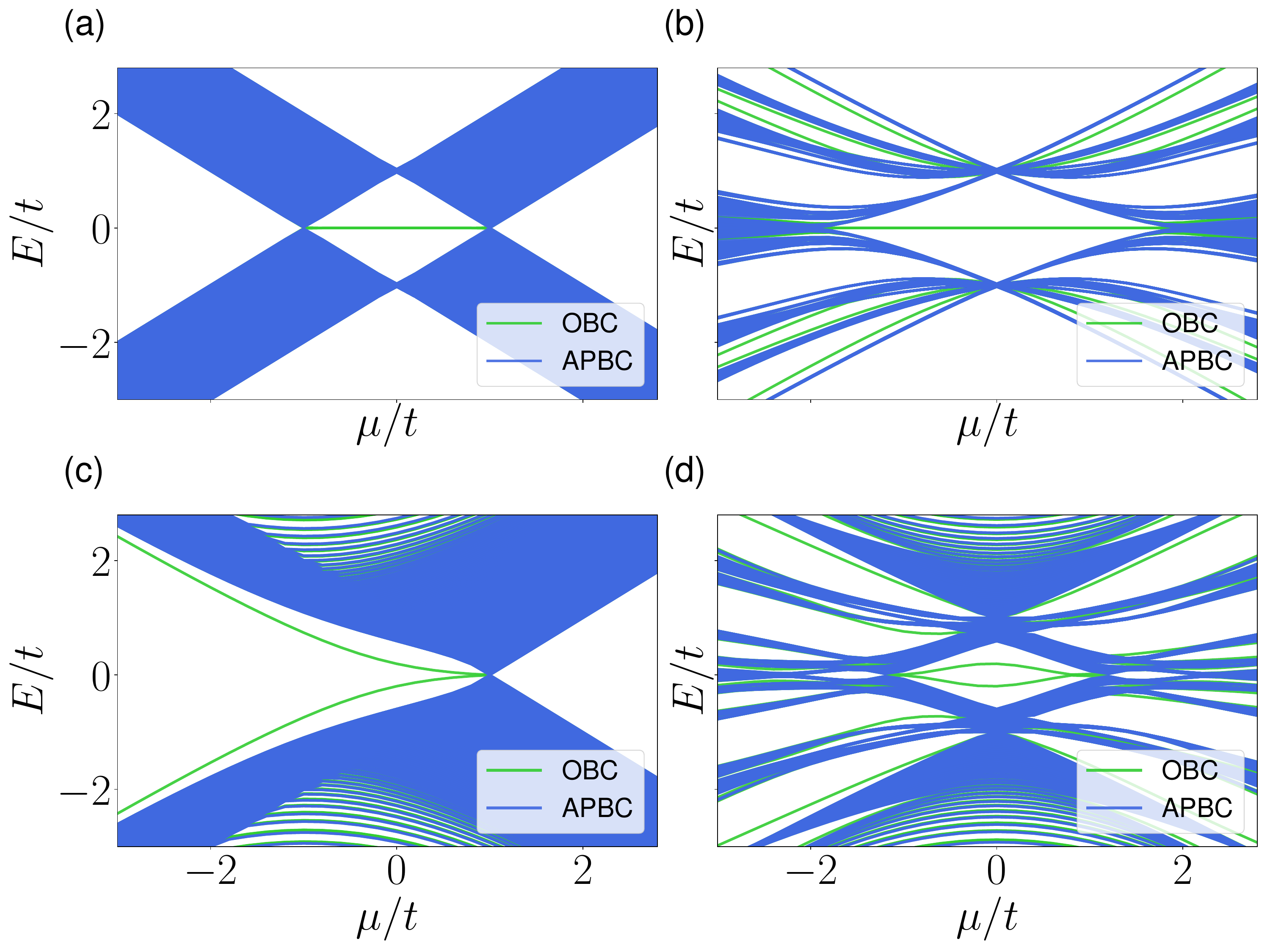}
\caption{\label{fig:bands} Energy spectrum for the system with OBC (green) and APBC (blue) for $q=233$, $L=1$, $\Delta = 0.5t$ and $\phi = 0$. We consider $f(i) = 1$ in (a) and (c) and AAH chemical potential in (b) and (d). Panels (a) and (b) depict the energy spectrum for the short-range system ($\alpha=5.0$) and thus, exhibit MMs. Panels (c) and (d) depict the energy spectrum for the long-range system ($\alpha=0.5$), which harbours MDMs in the central gap.}
\end{figure}

We first study the energy spectrum as a function of the chemical potential $\mu$, comparing the case of homogeneous chemical potential, $f(i)=\text{cst}=1$, with the case of a modulated AAH potential. We consider a system of 233 sites, and for the AAH potential, we take the modulation frequency $\beta$ to be $p/q = 144/233$. We expect, however, that the  results qualitatively hold also for other values of $\beta$, if $q$ is sufficiently large.

Figures~\ref{fig:bands} (a) and (c) depict the energy spectrum of the system for homogeneous chemical potential with APBC (blue) and OBC (green). For the short-range system (a), we see the appearance of zero-energy MMs in the central energy gap for $\vert \mu/t \vert < 1$. For $\vert \mu/t \vert > 1$, the system is topologically trivial, and no edge states exist within the bandgap. For the long-range system, shown in Figure~\ref{fig:bands}~(c), we only see one gap closing at $\mu/t = 1$. For $\mu/t < 1$, the system exhibits MDMs, which are clearly separated from the bulk and only exist for OBC. For $\mu/t > 1$, in contrast, the system is topologically trivial and does not exhibit any edge states. Figures~\ref{fig:bands} (b) and (d) depict the energy spectrum for a system with an AAH potential. The AAH potential leads to a splitting of the spectrum into many bands. Band gaps at non-zero energy and the massive edge states hosted by these gaps  will be analyzed in Section~\ref{sec:level5}. In the present Section, we focus on the central gap around $E=0$. For the short-range system, shown in Figure~\ref{fig:bands}(b), we observe the existence of the zero-energy MMs in the central region around $\mu \approx 0$. As compared to the case of homogeneous potential, the AAH modulation significantly increases the parameter range supporting MMs, which now extends to $\vert \mu/t \vert \lessapprox 2$.  For the long-range system, shown in Figure~\ref{fig:bands}(d), the MMs  hybridize to form MDMs. In this case, the central region for which MDMs exist is $\vert \mu/t \vert < 1$. Interestingly, the MDMs exhibit a crossing at zero energy which will be further discussed in Sec. \ref{sub:level45}. Moreover, at $|\mu/t|\approx 1.5$, we see another opening of weakly gapped regions which do not host MDMs or MMs. 

We conclude that the Kitaev chain with AAH potential exhibits a variety of different phases which are distinguished through the presence or absence of MMs and MDMs. In the following, it will be our goal to further characterize these different phases through the winding numbers introduced in Sec. \ref{sec:level3}.

\subsection{\label{sub:level42} Comparison between real-space and infinite system winding numbers}
\begin{figure}[t]
\includegraphics[width=1\columnwidth]{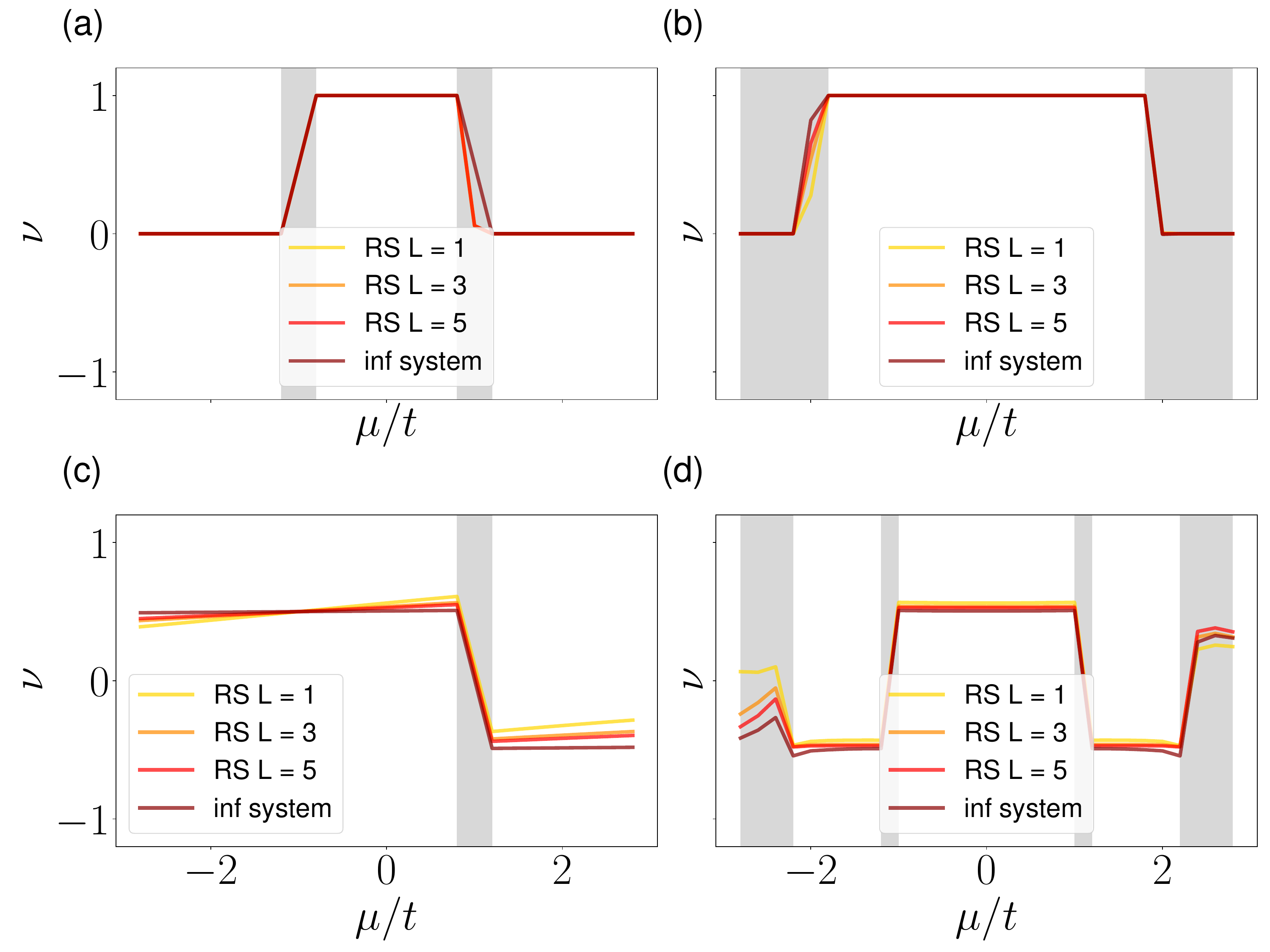}
\caption{\label{fig:wn} Real-space winding number for $L=\{1,3,5\}$ and infinite system winding number. In both cases, we considered $q = 233$, $\Delta = 0.5t$ and $\phi=0$. We consider $\alpha=5.0$ and $f(i)=1$ in (a), $\alpha=5.0$ and AAH chemical potential in (b), $\alpha=0.5$ and $f(i)=1$ in (c) and $\alpha=0.5$ and AAH chemical potential in (d). We see that, when increasing $L$, the real-space winding number converges to the one for the infinite system. The convergence is slower for long-range systems (c-d).  Note that the computation of winding number is not valid for the values of $\mu$ for which there is no gap at zero energy (shaded regions). }
\end{figure}

Before performing an in-depth analysis of the winding numbers and the edge states of the Kitaev chain in the presence of an AAH potential, we compare the two algorithms to compute the winding number introduced in the previous Section. Figure~\ref{fig:wn} shows the comparison between the real-space winding number for different values of $L$ and the infinite system winding number. For the short-range system (a)-(b), where we consider $\alpha=5$, the winding number takes the discrete values $0$ and $1$. Note that for the short-range case, the FHS algorithm from Section \ref{sub:level31} would lead to valid results. The non-trivial central region for which $\nu \neq 0$, which is $\vert\mu/t\vert < 1$ for $f(i)=\text{cst}$ (a) and $\vert\mu/t\vert < 2$ for AAH onsite potential (b) coincides with the existence of MMs in Figures~\ref{fig:bands}(a)-(b). The real-space winding number here is computed for a system with $L=\{1,3,5\}$, while the infinite system winding number takes $L\rightarrow \infty$. For the short-range system, both algorithms show good qualitative agreement even for $L=1$. Thus, here $N=q=233$ is large enough for the approximation in Eq.~\eqref{eq:approx_commutator} to converge.

Figures~\ref{fig:wn} (c)-(d) depict the winding number for the long-range system with $\alpha = 0.5$. Here again, the real-space winding number is computed for a system with $q=233$ and $L=\{1,3,5\}$. The winding number takes half-integer values, $\pm 1/2$. In the case $f(i)=cst$, shown in panel (c), the positive winding number corresponds to the region for which $\mu/t < 1$, which is also the region that corresponds to the presence of MDMs [See Figure~\ref{fig:bands}~(c)]. For $\mu/t>1$, the system does not exhibit any edge states but it is neither trivial, since the winding number takes a non-zero value: this system exhibits weak bulk-edge correspondence~\cite{lepori17}. For the AAH onsite potential, shown in Figure~\ref{fig:wn}~(d), the positive winding number corresponds to the region where $\vert \mu/t \vert < 1$. Again this is the region for which the system presents MDMs in the energy spectrum, cf. Figure~\ref{fig:bands}(d).  The two neighboring regions, $1.5 < \vert\mu/t\vert <2$, show a negative value of the winding number and do not exhibit MDMs. Note that the gap closes for $\vert\mu/t\vert > 2$, and correspondingly, the winding number is no longer well defined, as expected.

For the long-range case, the real-space winding number does not agree exactly with the infinite winding number for the values of $L$ that we have considered. The reason for this is that for long-range systems, the approximation in Eq.~\eqref{eq:approx_commutator} converges more slowly, which leads to small finite-size effects at $L=1$.

\subsection{\label{sub:level43} Critical superconducting pairing in short-ranged and long-ranged systems}

\begin{figure}[t]
\includegraphics[width=\columnwidth]{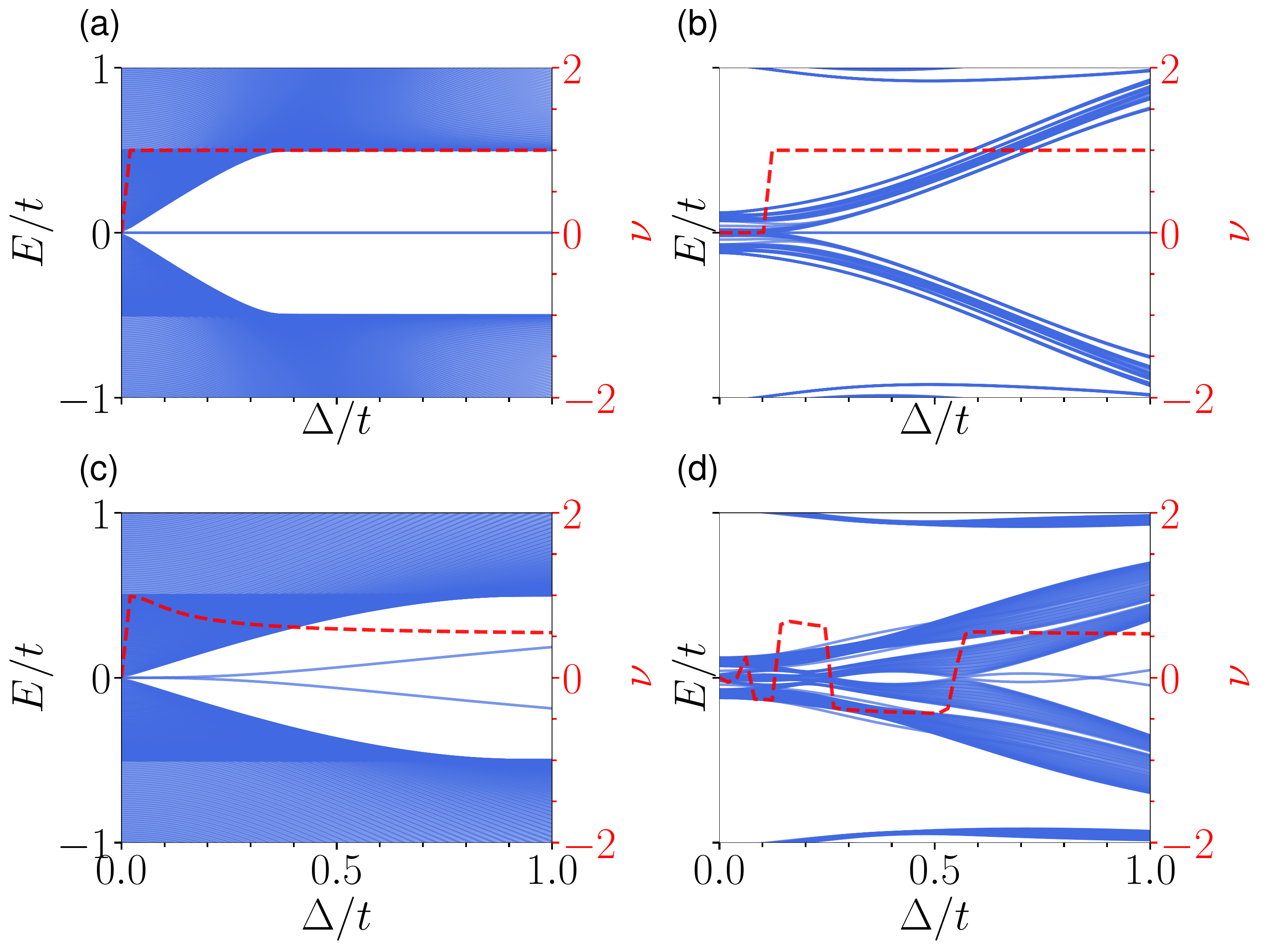}
\caption{\label{fig:critical_delta} Energy spectrum (blue) and winding number (red) for $f(i) = 1$, $\alpha=5.0$ and $\mu/t = 0.5$ (a), AAH chemical potential, $\alpha=5.0$ and $\mu/t = 1.2$ (b), $f(i) = 1$, $\alpha=0.5$ and $\mu/t = 0.5$ (c) and AAH chemical potential, $\alpha=0.5$ and $\mu/t = 1.2$ (d). In (a) and (c), the system exhibits edge modes for any $\Delta_C > 0$. For short-range pairing, the edge states are zero-energy modes and the winding number is equal to one, see panel (a). For long-range pairing, the edge states become massive, and the winding number becomes a half-integer, see panel (c). In contrast, in presence of an AAH potential, there is a region $\Delta < \Delta_C$ for which the central gap is closed, for both short-range pairing [panel (b)] and long-range pairing [panel (d)]. In the latter case, the combined effect of AAH onsite potential and long-range pairing leads to a succession of topologically differentiated regions separated by gap closings, see panel (d). As for the case of constant potential, the winding numbers in presence of AAH potential are integer in the short-ranged system, and half-integer in the long-ranged system. All quantities are computed using the real-space Hamiltonian for a system with $q=233$ and $L=1$.}
\end{figure}

We now investigate how the presence of MMs and MDMs depends on the value of the superconducting pairing $\Delta$. To this end, we compute the energy spectra of the finite Hamiltonian with OBC as a function of $\Delta$, as well as the real space winding number for the same system size with APBC. The results are depicted in Figure~\ref{fig:critical_delta}, showing energies in blue and winding number in red. 

Figures~\ref{fig:critical_delta}~(a)-(b) show the results for a system with rapidly decaying pairing ($\alpha =5$). Panel (a) considers a system with homogeneous chemical potential ($\mu/t=0.5$) and panel (b) considers a system with AAH chemical potential ($\mu/t=1.2$). For the homogeneous potential, the system exhibits topological edge modes even in the limit $\Delta \rightarrow 0$. In contrast, the presence of the AAH potential leads to a critical pairing $\Delta_C>0$ which is required for the opening of the gap and the appearance of MMs.  The real space winding number also becomes non-zero only for $\Delta >\Delta_C$.  This consistency between number of edge modes and topological invariant is an example of the celebrated bulk-edge correspondence.

Figures~\ref{fig:critical_delta}~(c)-(d) depict energy spectrum and winding number for a slow power law decay ($\alpha=0.5$), again distinguishing between homogeneous potential (c), and AAH potential (d). The critical pairing also remains $\Delta_C=0$ for homogeneous chemical potentials. However, the combination of an AAH potential and long-range pairing is found to yield a very rich scenario:  the system is gapless for pairing terms up to $\Delta=0.25t$. At this value, a gap opens and closes at $\Delta_C=0.55t$. There are no edge states within this gap. For $\Delta>0.55t$, another gap opens, but now containing MDMs. As already seen in the energy spectrum as a function of $\mu$, the MDMs cross at zero energy at $\Delta=0.85t$. A discussion of this interesting behavior will be given in Section \ref{sub:level45}. The gapped regimes are characterized by non-zero winding numbers: The winding number takes the value $\nu=-0.5$ in the first gap (the one without edge states), and $\nu=0.5$ in the second gap (the one with MDMs). Thus, the appearance of MDMs is accompanied by a jump of the winding number by 1, but we emphasize that in general, for $\alpha<1$, the system has a weak bulk-boundary correspondence~\cite{lepori17}. This means that a non-zero value of the winding number does not always correspond to MDMs. In Appendix~\ref{appendix:WBBC}, we show that the bulk-boundary correspondence is weak by looking at the system when varying the superconducting pairing $\Delta$ from negative to positive values. In this case, the winding number also jumps from $-0.5$ to $0.5$, but the edge physics remains the same. Nevertheless, the value of the winding number has physical consequences in the distribution of the Schmidt eigenvalues and the violation of the area law for the von-Neumann entropy~\cite{lepori17} or the quantization or half quantization of the multipartite entanglement~\cite{pezze17}.

\subsection{\label{sub:level44} Phase diagram}

\begin{figure}[t]
\includegraphics[width=\columnwidth]{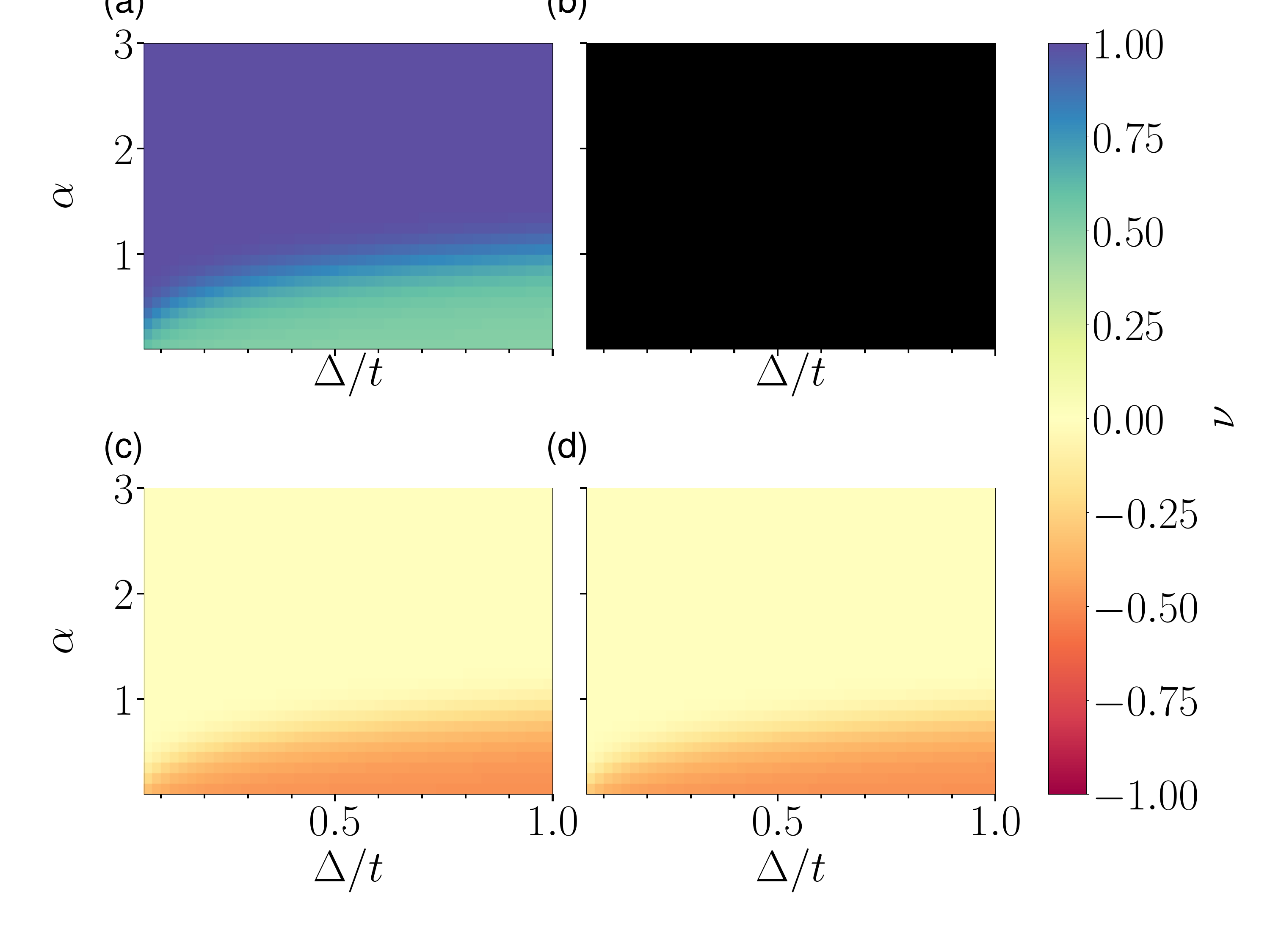}
\caption{\label{fig: cd_constant} Real-space winding number for the system with $f(i)=1$ and $\mu/t \in \{0.5,1,1.5,2\}$ (a-d). We observe two regions, corresponding to $\alpha<1$ (long-range) and $\alpha>1$ (short-range). At $\mu/t = 1$ (b) there is a closing of the gap, for which the winding number is not well-defined. When we go from $\mu/t<1$ (a) to $\mu/t>1$ (c-d) we see a topological phase transition, in which the winding number for short-range goes from $1$ to $0$ and for long-range from $0.5$ to $-0.5$. Here, $\Delta_C=0$. The real-space winding number is computed for $q=233$ and $L=1$. We consider that the central gap is closed when $\Delta E < 0.035t$ (black regions).}
\end{figure}

\begin{figure}[t]
\includegraphics[width=\columnwidth]{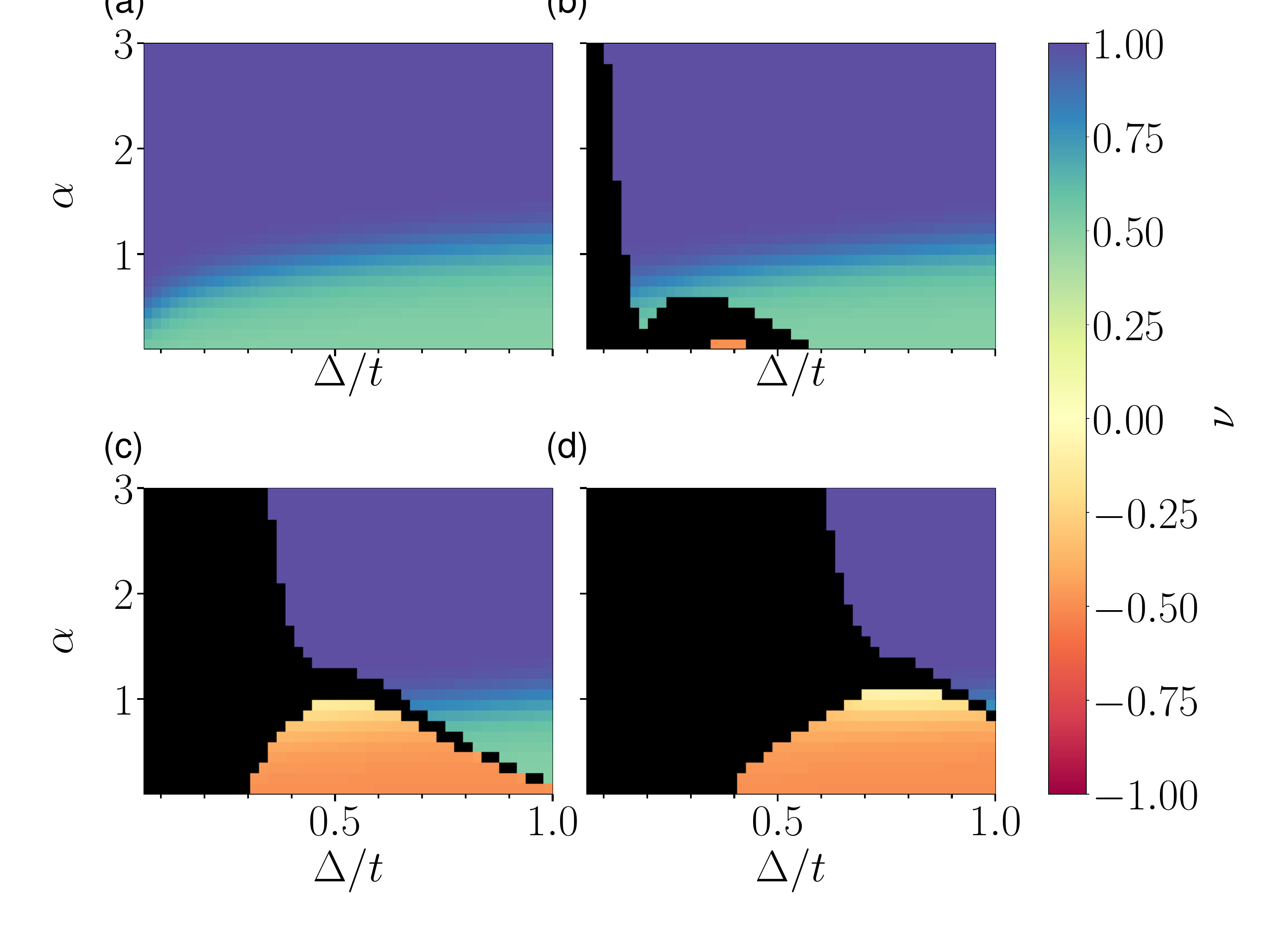}
\caption{\label{fig: cd_noconstant} Real-space winding number for the system with AAH onsite potential and $\mu/t \in \{0.5,1,1.5,2\}$ (a-d). When $\mu/t < 1$ (a), we only observe two regions, corresponding to $\alpha<1$ (long-range) and $\alpha>1$ (short-range). Here, the system behaves like in the homogeneous case in Figure~\ref{fig: cd_constant}, and thus $\Delta_C = 0$. When $\mu/t\geq 1$, the critical superconducting pairing $\Delta_C \neq 0$ both for the short-range and the long-range cases. For the short range case there is simply a region where the gap is closed for $\Delta<\Delta_C$ and a region with winding number $1$ for $\Delta > \Delta_C$. For the long range case, we see three different regions: first the gap is closed, then it opens with winding number $-0.5$, then it closes again and it re-opens with winding number $0.5$. The real-space winding number is obtained for a system with size $q=233$ and $L=1$. We consider that the central gap is closed when $\Delta E < 0.025t$ (black regions).}
\end{figure}

We now study the phase diagram of the model by fixing the chemical potential to a discrete value $\mu/t \in \{0.5,1,1.5,2\}$, while continuously varying $\alpha$ and $\Delta$. The phase diagrams are expressed in terms of the winding number, shown in Figure~\ref{fig: cd_constant} for homogeneous chemical potentials, and in Figure~\ref{fig: cd_noconstant} for AAH potentials.

For the homogeneous chemical potential, 
Figure~\ref{fig: cd_constant}, we observe that the critical superconducting pairing is always at $\Delta_C=0$. We see two regions, one corresponding to the short-range Kitaev chain ($\alpha > 1$), and one corresponding to long-range system ($\alpha<1$). In the short-range case, the winding number is $1$ for $\mu/t<1$, and $0$ for $\mu/t>1$. The long-range scenario has a winding number of $0.5$ for $\mu/t<1$, and $-0.5$ for $\mu/t>1$. At $\mu/t = 1$, the system is gapless and the winding number is ill-defined. We also observe in Figure~\ref{fig: cd_constant} (a) that, for small values of $\Delta$ and for $0.5<\alpha<1$, the long-range system exhibits MMs, which later hybridizes to form MDMs \cite{viyuela16} when $\Delta$ is increased.

For the AAH chemical potential, Figure~\ref{fig: cd_noconstant}, we make the following observations: For $\mu/t<1$ (a), the system behaves as in the homogeneous case, exhibiting two different regions, corresponding to long-range and short-range behavior. From this we can assume that, for very small chemical potential, the AAH modulation does not affect the topology of the system. However, for $\mu/t\geq 1$, the situation changes, see Figures~\ref{fig: cd_noconstant}~(b)-(d): the critical $\Delta_C$ is now shifted to a non-zero value. The latter seems to increase continuously for decreasing values of $\alpha$. For $\Delta< \Delta_C$, the central gap of the system is closed and the winding number is not defined. If we now focus on the region $\alpha<1$, we clearly see a region in which the sign of the winding number changes. This phase does not host MDMs and it is separated from the other phases through a gap closing. Finally, the value of $\Delta_C$ depends drastically on the value of the chemical potential amplitude $\mu$.

\subsection{\label{sub:level45} Edge modes in the slow power law decay regime}

\begin{figure}[t]
\includegraphics[width=\columnwidth]{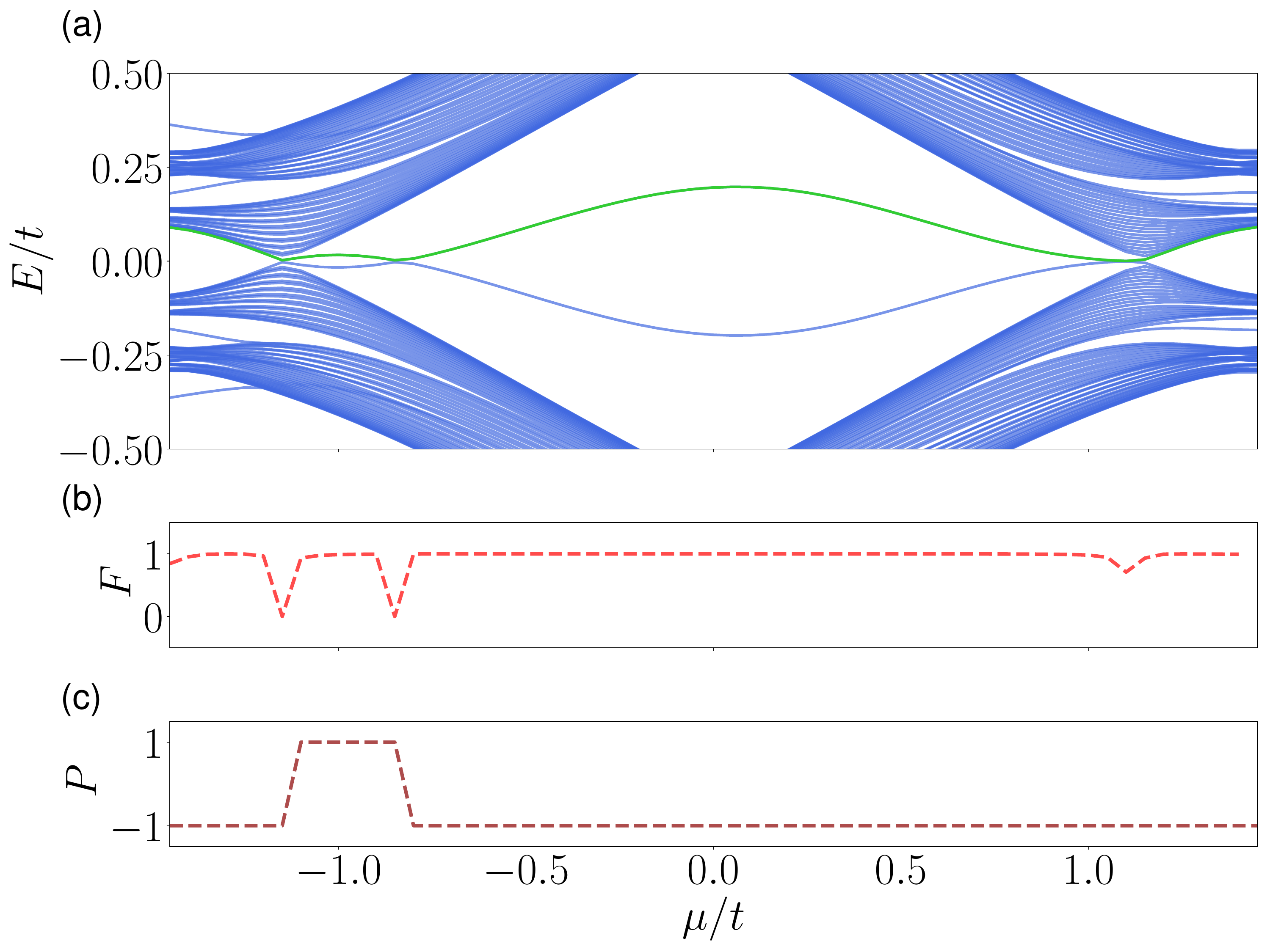}
\caption{\label{fig:crossing} Energy spectrum (a), quantum fidelity $F$ (b) and ground state fermion parity $P$ (c) for a system with size $q=233$ and $L=1$. Here we considered $\Delta = 0.5t$, $\alpha=0.5$ and $\phi = \pi$. The quantum fidelity $F$, which is computed following the MDM indicated in green, takes the value $1$ everywhere except for the closings of the gap at $\vert \mu/t \vert = 1.2$ and the crossing at $\mu/t = -0.85$. The ground state fermion parity switches at the gap closing at $\mu/t = -1.2$ and at the crossing at $\mu/t = -0.85$.}
\end{figure}

As observed in the previous subsections, the central gap of the (sufficiently) long-range Kitaev chain with AAH potential hosts an edge mode at positive energy and an edge mode at negative energy, but very interestingly, these modes exhibit crossings at zero energy [see Figure~\ref{fig:bands}~(d) and Figure~\ref{fig:critical_delta}~(d)]. In the present subsection, we focus on this crossing. 

First, we note that the chiral symmetry of the system dictates that a crossing of the two modes can only happen at zero energy. However, we are not aware of a symmetry which would prevent the two modes from mixing, and thus, they could also perform an avoided crossing. To distinguish a true crossing from an avoided one, we study the quantum fidelity between eigenstates as we adiabatically change the chemical potential. The fidelity is defined as
\begin{equation}
    F(\mu) = \vert \langle \psi_{\mu} \vert \psi_{\mu+\epsilon} \rangle \vert,
\end{equation}
where $\psi_{\mu}$ is the single-particle wave function describing the edge mode with positive energy. 

In \cite{hedge16}, they analyze the ground state fermion parity of the short-range Kitaev chain, showing that for finite systems this observable switches for certain values of the chemical potential. This finite size effect is a result of the energy splitting between the MM, which are not exactly at zero energy and undergo crossings for certain values of $\mu$. In the long-range scenario the energy splitting between the MDM remains finite even when the system size goes to infinity, but if we have a true crossing, we might expect a switch of the fermion parity at the crossing point. 

Following \cite{kitaev01}, the ground state fermion parity $P$ is defined as
\begin{equation}
    P = \text{sgn}(\text{Pf}(H_M)),
\end{equation}
where $\text{Pf}$ stands for Pfaffian and $H_M$ is the Hamiltonian in the Majorana basis with OBC. 

In Figure~\ref{fig:crossing},  we show the energy spectrum together with the quantum fidelity and the ground state fermion parity for a finite size system with OBC and $\alpha=0.5$.
 The crossing happens at $\mu/t=-0.85$. At this crossing point, the quantum fidelity drops to zero and the fermion parity switches its sign. This demonstrates that, at the crossing point, the two orthogonal modes are exchanged, and thus, that the crossing is indeed a true crossing and not an avoided one.
 
 Moreover, we also compute the winding number to see if the crossing corresponds to a change in the bulk topology. If one interprets the vicinity of the zero-energy crossing as a reappearance of Majorana modes, one would expect that the winding number becomes $1$. However, the crossing is found to have no effect on the winding number, which remains to be $0.5$ throughout the whole gapped region (i.e. for $\vert \mu/t \vert < 1.2$). Since the zero-energy modes only exist at the exact crossing point, a switch in the bulk topology might be difficult to capture. On the other hand, as mentioned already in Sec. \ref{sub:level43}, long-range interactions can weaken or destroy the bulk-edge correspondence, and therefore we cannot discard the posibility that the bulk topology does not match the edge dynamics.

\begin{figure}[t]
\includegraphics[width=1\columnwidth]{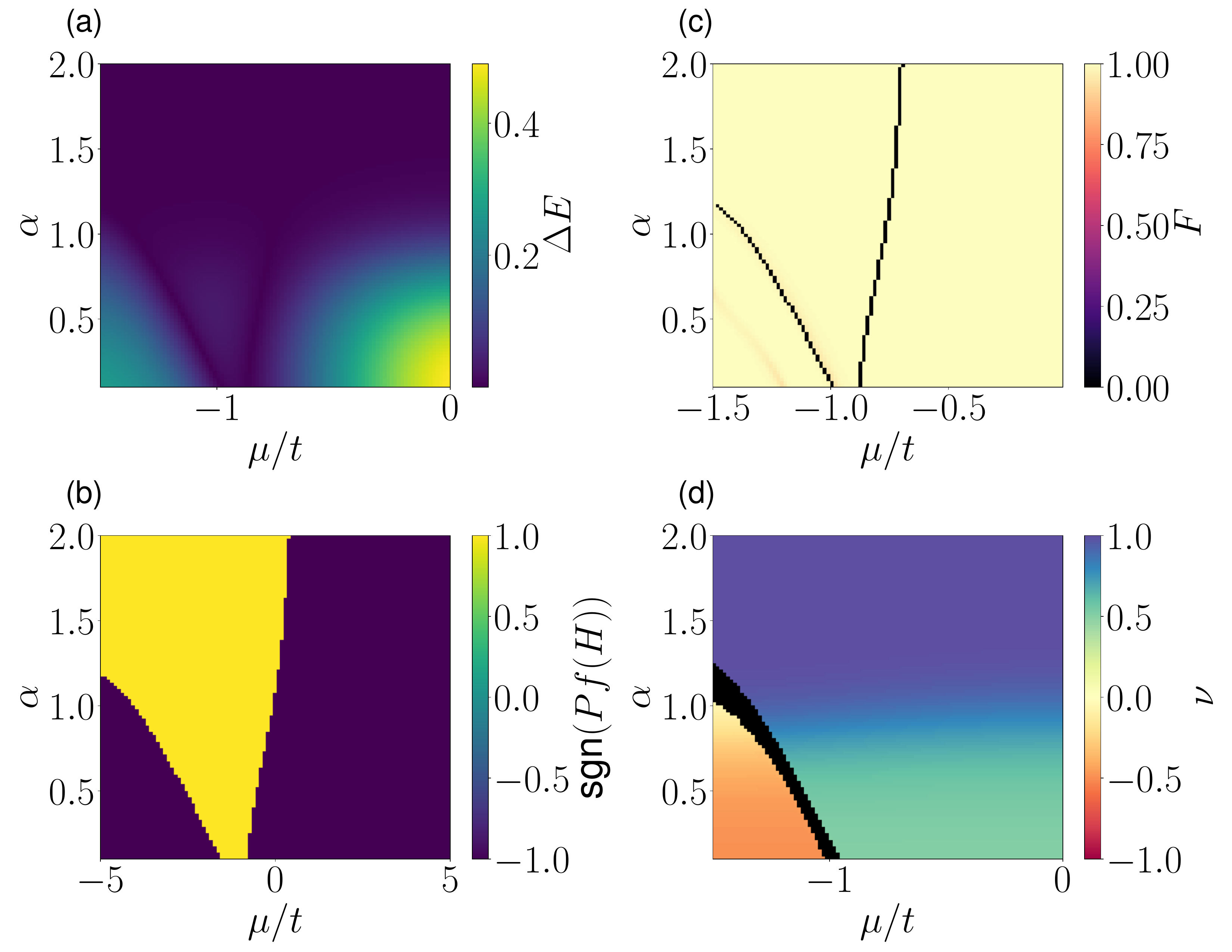}
\caption{\label{fig:crossing_cm}  Energy splitting $\Delta E$ of edge modes (a), quantum fidelity (b), ground state fermion parity (c) and real-space winding number (d) for different values of $\mu$ and $\alpha$. The energy splitting $\Delta E$, the quantum fidelity $F$ and the ground state fermion parity switches allow us to detect the crossings, while the winding number $\nu$ does not detect it. All the computations consider a system with $q=233, L=1$, and $\phi = \pi$. The winding number is computed using the real-space approach. Gapless regions (defined by a threshold $\Delta E < 0.02t$), where the winding number becomes ill-defined, are colored black in panel (c). Note that in (b) and (c) the region that corresponds to $\alpha>1$ is expected to host MMs, but even in this regime, there is a finite-size splitting of the two modes and a true energy crossing at zero energy occurs.
}
\end{figure}

For a better understanding of the relation between Majorana physics and the observed zero-energy crossings, we shall ask how the zero-energy crossing connects to the regime in which pairing is sufficiently short-range to support an extended phase with zero-energy modes. To this end, we plot in Figure~\ref{fig:crossing_cm} (a-d) the energy splitting $\Delta E$ between the two edge modes, the quantum fidelity $F$ of the positive mode, the ground state fermion parity $P$ and the winding number $\nu$ as a function of both $\mu$ and $\alpha$. For $\alpha>1$, the energy splitting becomes very small, and, as we demonstrate below, in this regime the finite value of the splitting is a mere finite-size effect. Therefore, the edge modes in this regime can be interpreted as Majorana zero modes, and such interpretation is corroborated by the winding number which takes the value $\nu=1$ for $\alpha>1$. Nevertheless, the finite-size splitting still allows to energetically differentiate between the two modes, and also for $\alpha>1$ we observe a true level crossing, as indicated by the black line of zero fidelity in Figure~\ref{fig:crossing_cm} (b) and the fermion parity switch in Figure~\ref{fig:crossing_cm} (c). Interestingly, this curve seamlessly extends to the level crossing at $\alpha<1$. In this regime, however, the splitting between the two modes takes significantly larger values [see Figure~\ref{fig:crossing_cm} (a)], and cannot any longer be interpreted as a finite-size effect. This demonstrates that the isolated zero-energy modes at $\alpha<1$ are smoothly connected to the MMs at $\alpha>1$. In contrast to this, the bulk topology is different in the two regimes: the winding number drops from 1 to 0.5 when $\alpha<1$  [see Figure~\ref{fig:crossing_cm} (d)]. We also mention that, for $\alpha<1$ and $\mu/t<-1$, the central gap closes [marked in black colors in Figure~\ref{fig:crossing_cm} (d) and yielding the second black line in Figure  ~\ref{fig:crossing_cm} (b) and a femion parity switch in Figure  ~\ref{fig:crossing_cm} (c)]. As can be seen from Figure~\ref{fig:crossing_cm}(b), the zero-energy crossing of the edge modes (right black line) and the gap closing (left black line) approach each other, when $\alpha$ is reduced. For completeness, we also mention that the gap re-opens at even smaller value of $\mu$. In this regime, no edge modes are observed, yet the winding number takes a non-zero value, $\nu = -0.5$. 

\begin{figure}[t]
\includegraphics[width=1\columnwidth]{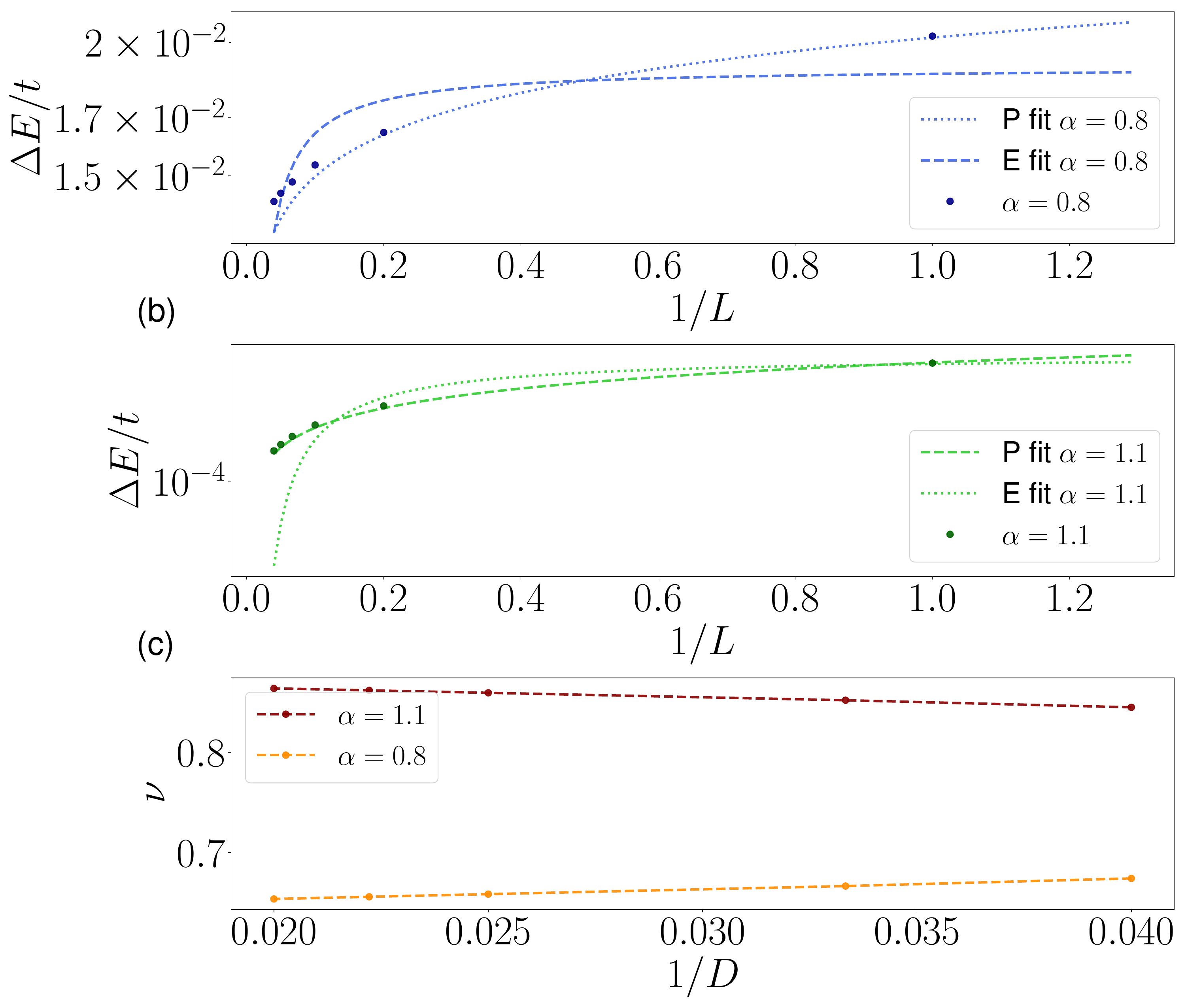}
\caption{\label{fig:scaling_alpha} Finite-size scaling of the energy splitting $\Delta E$ of the edge modes vs. $1/L$, where $L$ is the number of supercells, each containing $q=233$ sites. We choose $\mu/t = -1$, $\Delta = 0.5t$, $\phi = \pi$, so that we are on the left side of the crossing in Fig. \ref{fig:crossing}. We plot the data for $\alpha=0.8$ (a) and $\alpha=1.1$ (b), and compare both cases with exponential fits (dashed lines) and polynomial fits (dotted lines). For both fits, we have assumed that $\Delta E \rightarrow 0$ for $L\rightarrow \infty$ (see text). Discrepancy with both fits at $\alpha=0.8$ suggest that the splitting $\Delta E$ remains finite in this case.
(c) Scaling of the infinite system winding number with the discretization $D$ defining the computational grid of $\delta_k =\frac{2\pi}{D}$.}
\end{figure}

To conclude this section, we support our claim regarding the qualitative changes at $\alpha=1$ by a finite-size scaling of the degeneracy splitting of the modes, and of the winding number. Therefore, we focus on two values of $\alpha=0.8$ and $\alpha=1.1$, and fix $\mu/t=-1$ to a value on the left side of the zero-energy crossing. For this choice, we plot the mode splitting $\Delta E$ against the inverse of the system size, $1/L$, in Figure~\ref{fig:scaling_alpha} (a-b). For the finite-size scaling, we assume an exponential closing of the splitting, $\Delta E = a e^{-b L}$, or a polynomial closing of the splitting, $\Delta E= a (1/L)^b$, in both cases with two fit parameters $a$ and $b$. Because of the long-range character of the Hamiltonian, the dispersion relation contains now the HLP function, which depends on the length of the system $L$ \cite{vodola14}. As a result, the exponential fit matches the data neither at $\alpha>1$ nor at $\alpha<1$. In contrast, the polynomial fit perfectly matches  the data for $\alpha>1$. However, for the largest system sizes considered ($L=25$ supercells), the observed energy splitting at $\alpha=0.8$ remains above the polynomial fit. This discrepancy becomes more pronounced and affects more data points for smaller values of $\alpha$. Thus, we conclude that the energy splitting remains finite for $\alpha<1$, while it vanishes polynomially with the system size for $\alpha>1$. 

The qualitative change at $\alpha=1$ is also backed by the behavior of the winding number: While sufficiently above $\alpha=1$, it takes the value one, the winding number is 0.5 when $\alpha$ is sufficiently below one. In the vicinity of $\alpha \approx 1$, the winding number takes intermediate values between 0.5 and 1, no matter which algorithm we use to compute the winding number. However, we find that in the real-space algorithm the intermediate values are due to the finite size of the system, whereas for the chiral algorithm (considering an infinite system) the finite discretization of the momentum grid is the reason for the intermediate values. Scaling the winding number vs. system size or discretization, we observe that the invariant approaches the value one, if $\alpha>1$, or the value 1/2, if $\alpha<1$. This is illustrated 
in Figure~\ref{fig:scaling_alpha} (c), for the scaling of the winding number vs. discretization $D$ at $\alpha=0.8$ and at $\alpha=1.1$ This observation suggests a sharp change in the bulk topology at (or in the vicinity of) $\alpha=1$.

\section{\label{sec:level5} Aubry-Andr\' e-Harper edge states}

\begin{figure}
\includegraphics[width=\columnwidth]{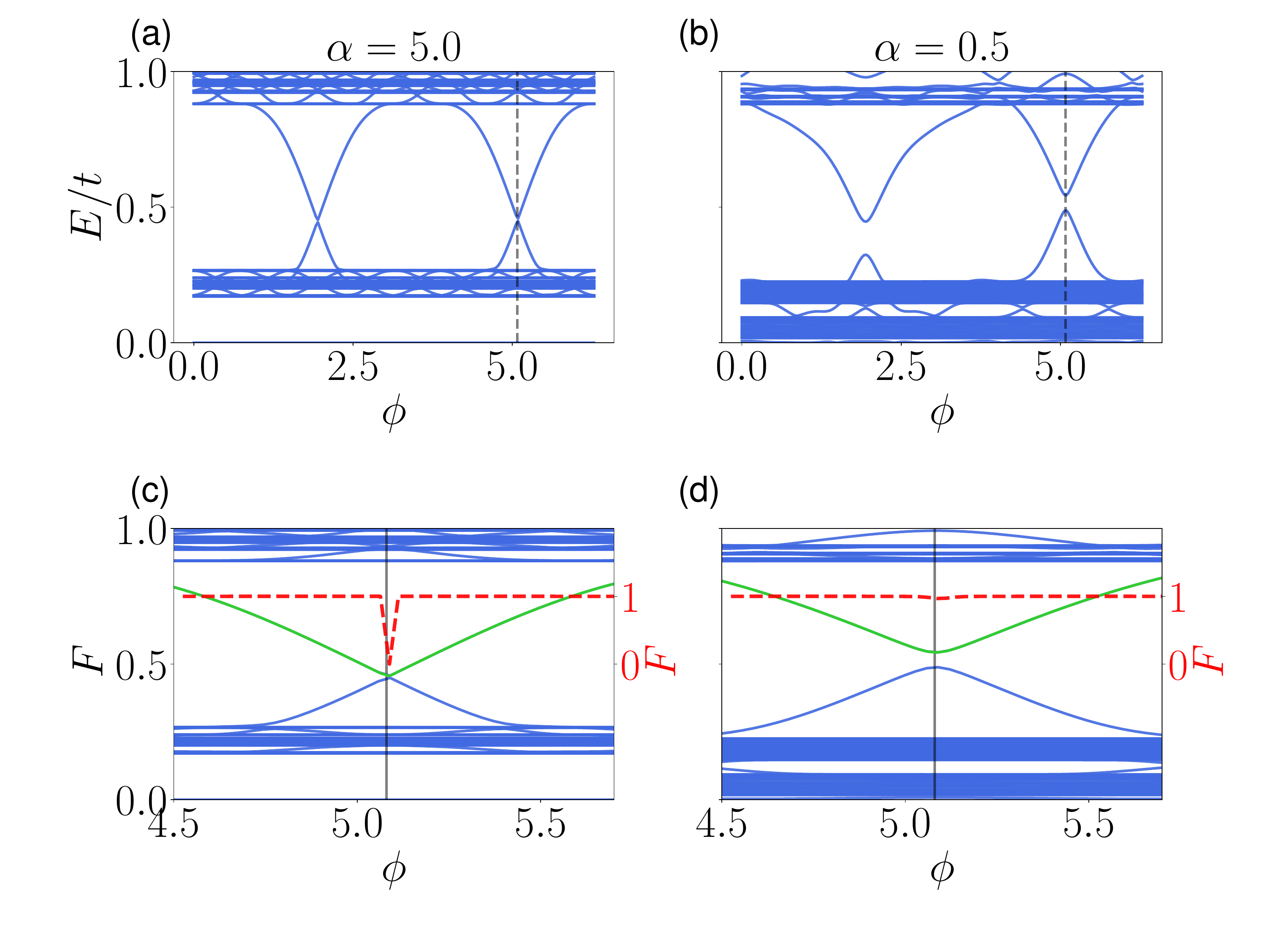}
\caption{\label{fig:AA_crossing} Energy spectrum as a function of $\phi \in [0,2\pi]$, in the vicinity of the energy gap around $E=0.5t$. Panels (a) and (c) correspond to the short-range system ($\alpha=5$). In (a) we see the energy spectrum for the full period of the dephasing, while in (c) we zoom into one of the avoided crossings. Here, we also compute the quantum fidelity $F$ (dashed red line). Panels (b) and (d) correspond to the long-range scenario ($\alpha=0.5$). Again, the upper panel (b) shows the energy spectrum for the full period of the dephasing, and the lower panel (d) zooms into the crossing and also considers the quantum fidelity $F$ (dashed red line). To compute the quantum fidelity, we follow the edge state highlighted in green. We considered a system with OBC and $q=233$, $L=1$, $\mu/t = 1$ and $\Delta = 0.3t$.}
\end{figure}

In the presence of an AAH potential, the system exhibits topological edge states in the higher energy gaps.  These AAH edge states arise from the AAH potential and can be understood through the dimensional reduction of the 2D Hofstadter model. The latter describes electrons in a 2D lattice subjected to a perpendicular magnetic field~\cite{hofstadter76} and its dimensional reduction onto a family of 1D chains corresponds to the AAH model. That is why the AAH edge states fall within the A class of the topological classification of 2D systems. In particular, different instances of the AAH model are characterized by the $2\pi$-periodic dephasing parameter $\phi$, which can be interpreted as one of the quasi-momenta of the original 2D system, i.e. $k_x = k$ and $k_y = \phi$. Historically, the pumping mechanism, i.e. relating the presence of edge states in an energy gap of 1D system through a 2D topological invariant, has been widely studied in systems containing an AAH potential \cite{kraus2012a}. Even though in the 1D system the AAH edge states are not topologically protected for any value of the dephasing parameter $\phi$, they are protected if there is a crossing point for which the symmetry of the system is higher. Recent studies such as \cite{yahyavi19} combine this mechanism with systems containing short-range superconducting pairing. In this Section, we study the fate of this mechanism for a system with long-range superconducting pairing.

In Figure~\ref{fig:AA_crossing}, we compare the edge states for the system with $\alpha<1$ (long-range) and with $\alpha>1$ (short-range) and focus on the energy gap around $E=0.5t$. In the normal (short-range) scenario [panel (a)], the edge states cross within the energy gap a given number of times. In the gap considered here, they cross twice (at $\phi=\{1.95, 5.07\}$). Figure~\ref{fig:AA_crossing}(c) shows the energy spectrum around the second crossing and analyzes the quantum fidelity $F$ at the crossing. It drops to zero at $\phi=5.07$. This must be contrasted to the situation in the long-range system. Figure~\ref{fig:AA_crossing}(b) depicts the energy spectrum for the long-range system with OBC, and two avoided crossings are observed at the same values of $\phi =  \{1.95, 5.07\}$. However, the quantum fidelity $F$ in Fig.~\ref{fig:AA_crossing}(d) stays at $1$ around $\phi = 5.07$. This indicates that the long-range superconducting pairing hybridizes the two edge states, such that the total charge transported in one period of the dephasing now becomes zero.

We now focus on the computation of the 2D Chern numbers. They characterize the bulk topology associated to these edge states, and for a system with bulk-edge correspondence, the value of the two-dimensional Chern number should coincide with number of crossings, since it indicates the charge that is pumped at the edges of a 1D cylinder. Here, we use the the Thouless-Kohmoto-Nightingale-Nijs (TKNN) algorithm introduced in Ref.~\cite{thouless82} where they define the Berry curvature as follows

\begin{equation}
    \Omega_{k_x k_y} = i\sum_{\substack{E_{\alpha}<E_F \\ E_{\beta}>E_F}} \frac{\partial_{k_x} H_{\alpha\beta}\partial_{k_y} H_{\beta\alpha}-\partial_{\partial k_y} H_{\alpha\beta}\partial_{k_x} H_{\beta\alpha}}{(E_{\alpha}-E_{\beta})^2},
\end{equation}
where $k_x$ and $k_y$ are quasi-momenta of the Brillouin zone. Recall that here, $k_x = k$ and $k_y = \phi$. We choose the value of the Fermi energy $E_F$ to be located inside the gap that we want to characterize. Note that
\begin{equation}
    \partial_{ k_x} H_{\alpha\beta} = \langle\psi_\alpha\vert \partial_{k_x} H  \vert \psi_\beta\rangle,
\end{equation}
where we consider the analytical derivative of the infinite Hamiltonian in Eq.~\eqref{eq: hamiltonian inf}. Then, the Chern number is defined as
\begin{equation}\label{eq: chern int}
C= \frac{1}{2\pi i}\int_\text{BZ} \Omega_{k_x k_y}.
\end{equation}
We here discretize the integral as a Riemann sum
\begin{equation}
    C = \frac{1}{2\pi i}\sum_{k_x,k_y}  \Omega_{k_x k_y} \Delta k_x \Delta k_y,
\end{equation}
where $\Delta_{k_x}$ and $\Delta_{k_y}$ are the two discretization lengths. In order to avoid the singularity that we mentioned in Section~\ref{sec:level3}, we consider a discretization of the Brillouin zone that does not explicitly contain the point $k=0$. Nevertheless, the integral in Eq.~\eqref{eq: chern int} is still well defined and converges to the expected value even if the singularity is included.

\begin{figure}
\includegraphics[width=\columnwidth]{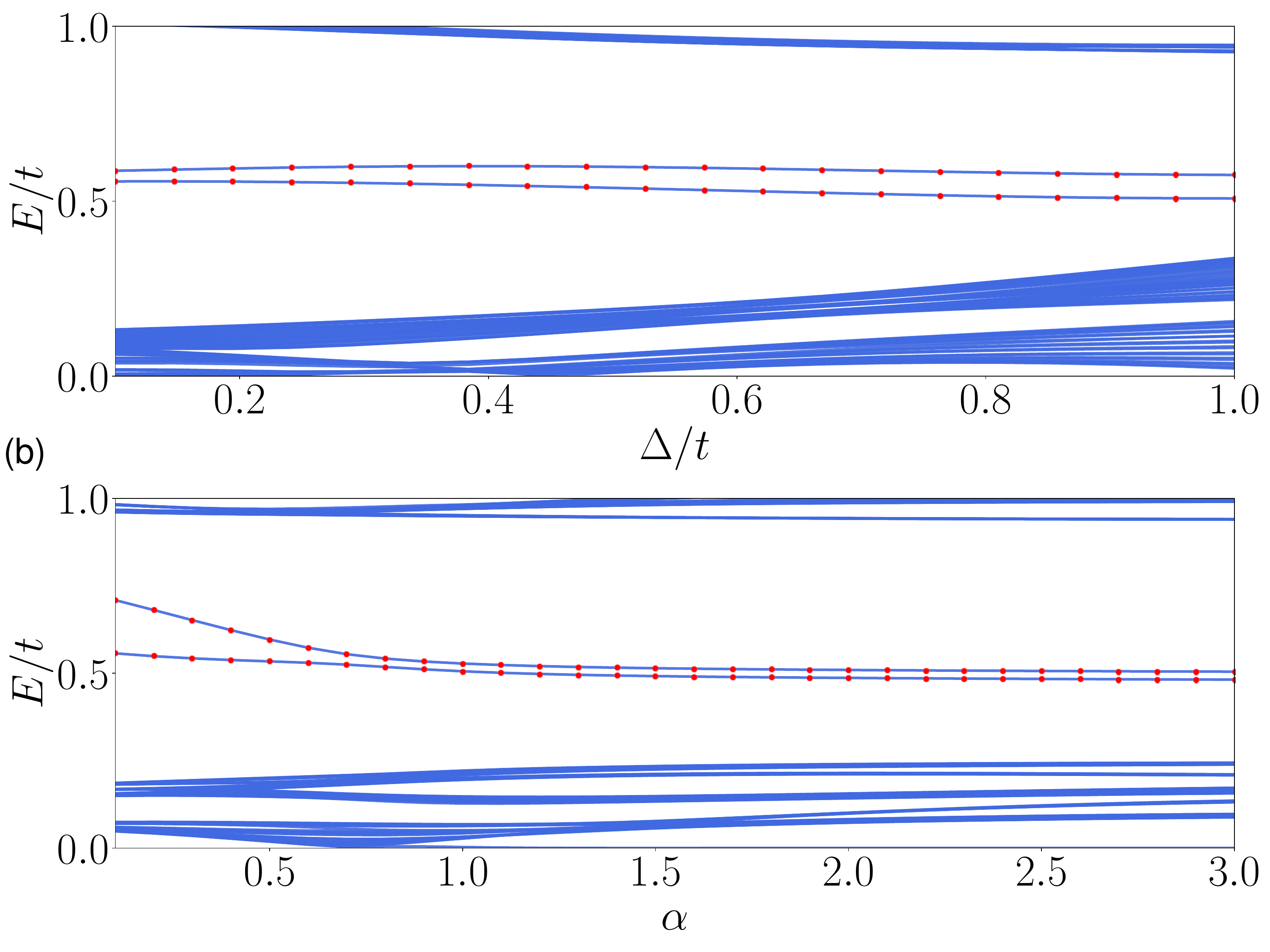}
\caption{\label{fig:AA_edge_states} (a) Energy of the AAH edge states vs. pairing strength $\Delta$ for fixed $\alpha = 0.5$. (b) Energy of the AAH edge states vs. decay exponent $\alpha$ for fixed $\Delta=0.3t$. In both plots, (a) and (b), we take the value of the phase at the crossing $\phi = 5.07$. In (a), the edge states split as we increase $\Delta$, while in (b) the edge states come together as we increase $\alpha$. The plots are obtained for a system with $q=233$, $L=1$ and $\mu/t=1.2$.}
\end{figure}

\begin{figure}
\includegraphics[width=\columnwidth]{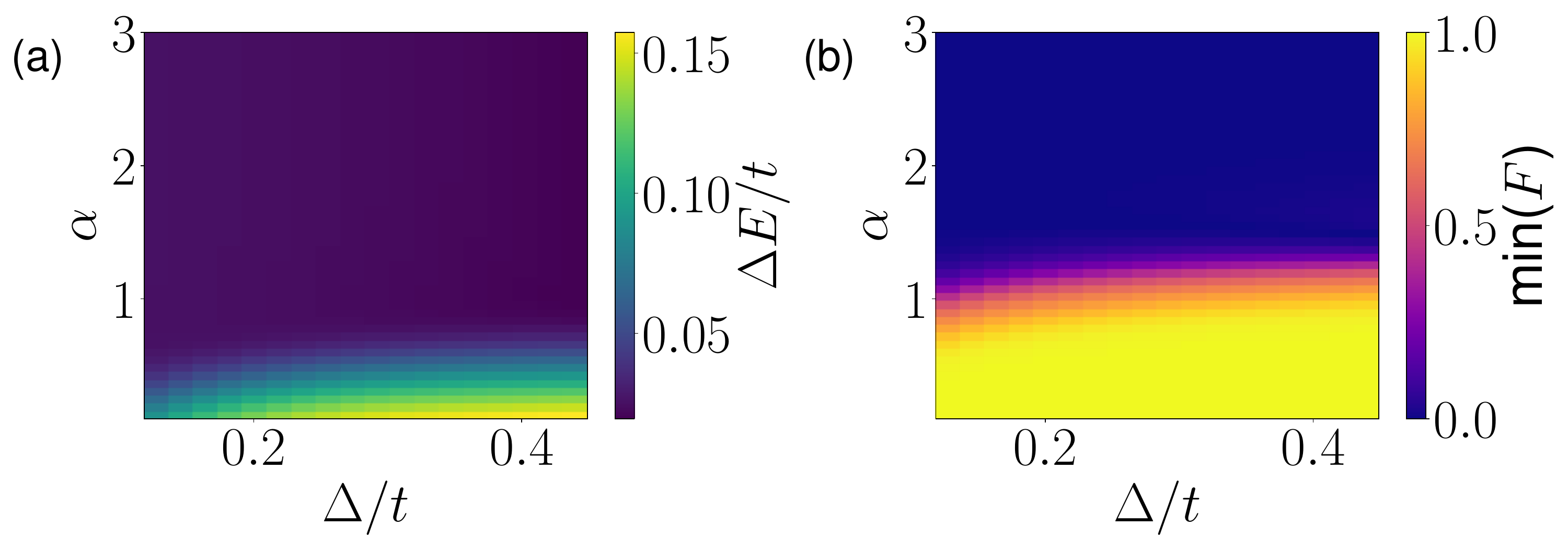}
\caption{\label{fig:AA_colormap} (a) Energy gap between the edge states at $\phi=5.07$, and (b) minimum of the quantum fidelity $F$ in the region $\phi\in [5,5.1]$. The plots are obtained for a system size of $q=233$, $L=1$ and $\mu/t=0.5$. They both indicate an avoided crossing of the AAH modes for the region where $\alpha<1$. }
\end{figure}

We see that at the energy gap around $E=0.5t$ the 2D Chern number takes the value $2$ for any value of the decay exponent $\alpha$ and of the superconducting pairing $\Delta$. This does not agree with the number of crossings within the gap, which we extract from Figures ~\ref{fig:AA_edge_states} and ~\ref{fig:AA_colormap}. In particular, in Figure~\ref{fig:AA_edge_states} we study the crossing at $\phi = 5.07$. We identify the edge states in the energy spectrum of the system with OBC and study them while changing the parameters of the Hamiltonian. We see that the crossing splits for $\alpha<1$ and increasing values of $\Delta$. In Figure~\ref{fig:AA_colormap} (a), we show the energy gap $\Delta E$ between the AAH edge states at the crossing point, which clearly increases for $\alpha<1$ and for any value of $\Delta$. Accordingly, in Figure~\ref{fig:AA_colormap} (b) the quantum fidelity $\text{min}(F)$ shows a minimum at zero for short-range systems but it does not drop to zero when the system is long-range, indicating that there is no crossing. 

We can conclude that the bulk topology arising from the AAH potential does not match the edge dynamics in the presence of the long-range superconducting pairing, and from this we can infer that the weakening of the bulk-boundary correspondence extends, not only to the MDMs but also to the AAH topology.

\section{\label{sec:level6} Experimental realization}

Before concluding we discuss shortly the experimental feasibility of realizing the 1D long-range Kitaev model  with Aubry-Andr\'e-Harper potential. Below we consider condensed matter systems and ultracold atomic systems. 

\subsection{\label{sub:level61} Condensed matter systems}

Although the quest for a Majorana fermion as a fundamental particle has been illusive, there has been a lot of experiments performed in solid state semiconductor/superconductor devices \cite{mourik12, rokhinson12, deng12, das12, churchill13, finck13, stanescu13, jia17, wang18, hart14, nadj14, deng16, zhang18, he17, xu15, sun16} aimed at creating Majorana fermions as quasiparticle excitations. Most of these experimental systems are 1D nanowires \cite{mourik12, rokhinson12, deng12, das12, churchill13, finck13, stanescu13, nadj14, deng16} , which are modeled fairly accurately by the  short-range Kitaev chain. These experimental setups either consist of: (a) a semiconducting nanowire with spin orbit coupling, placed on a conventional $s$-wave superconductor which provides proximity-induced superconductivity, while being subjected to a perpendicular magnetic field; (b) a chain of magnetic adatomic impurities (with frozen electronic motion) deposited on a superconducting surface, thus possessing local moments that induce Yu-Shiba-Rusinov bound states \cite{yu65, shiba68, rusinov69}  within the superconducting gap, and thereby forming an effective Kitaev chain \cite{wang18}. 

In case (a) under the strong coupling limit, it is theoretically assumed that the magnitude of the proximity-induced superconductivity is the same as that of the host superconductor. In experiments, however, the magnitude of the induced superconductivity is actually found to be much smaller. To take this into account, it is necessary to assume a weak-coupling between the surface system and the host superconductor, where it can be shown that exponentially decaying oscillating long-range pairing and hopping can arise in the  chain with a characteristic length scale of the order of the coherence length of the Cooper pairs. Hence, the coherence length is many times larger than the lattice constant, and the long-range pairings are practically non-local \cite{wang18}. 

In case (b) of a chain of magnetic impurities placed in a conventional superconductor, each individual adatom can create localized, individual sub-gap Shiba states. These states hybridize with each other and with the bulk superconducting condensate and, through multiple Andreev reflections, form energy bands. This band structure resembles that of the  Kitaev chain for a helical spin structure of the adatoms and can be probed by scanning tunneling spectroscopy. In the particular case of deep-lying Shiba states, it has been shown in Ref.~\cite{pientka13} that the effective tight-binding Bogoliubov-de Gennes Hamiltonian representing the  Kitaev chain has hoppings and pairings that are long-range, with a $1/r$-power-law decay within the coherence length of the host superconductor. Beyond this length scale, the decay is exponential. 

In both cases, the semiconducting nanowire in a magnetic field, and the nanowire-like chain of magnetic impurities are plagued by defects, either originating from the spatial variance of the magnetic field, or through the distribution of the impurities constituting the chain. This can be well modeled via the Aubrey-Andre model which includes the effect of spatial inhomogeneity. Recently, it has also been proposed that an effective  Kitaev chain with long-range pairing and hopping terms can also be formed from the combination of a solid-state Majorana platform made out of a planar Josephson junction in proximity to a 2D electron gas with Rashba spin-orbit coupling placed in an external magnetic field~\cite{balatsky06}. 

\subsection{\label{sub:level62} Utracold atoms}

We now discuss the building blocks necessary to realize such a system in a cold-atom quantum simulator. On the one hand, realizing a setup that generates the Aubry-Andre modulation is possible. In  fact, it has been demonstrated in numerous experiments using super lattice techniques, pioneered in Refs.~\cite{billy08, lahini09}. On the other hand, there has also been some proposals to realize the Kitaev chain in cold-atomic setups~\cite{kraus12a, buhler14}. The main difficulty for our Hamiltonian is to propose a cold-atomic setup that includes long-range power law superconducting pairing with a controlled decay rate. One possibility would be to engineer the long range pairing with the help of an attractive long-ranged interaction. Here we discuss three classes of cold atom setups that could lead to such interactions:

(i)\textit{Fermionic atoms/molecules with static dipole moments.}  Ultracold dipolar gases are in center of interests of the ultracold atoms quantum matter community since the end of 1990s (for reviews/books see Refs.~\cite{baranov08, lahaye09, lewenstein12, baranov12}). Static dipoles can be magnetic, as it happens in ``magnetic atoms'' like Chromium (condensed by T. Pfau group~\cite{griesmaie05}), Erbium (condensed by F. Ferlaino's group~\cite{aikawa12}), or Dysprossium (condensed by B. Lev's group~\cite{lu11}). Much stronger static electric dipoles can be induced in hetero-nuclear  molecules (first achieved in D. Jin's-J.Ye's groups~\cite{ni08}). Strong dipole-dipole interactions lead to non-standard Hubbard models, with many ingredients necessary for realization of the long-range Kitaev chain~\cite{dutta15}. The progress in  the field of ultracold dipolar gases, and in particular in  quantum engineering of chemistry/quantum matter with ultracold molecules is spectacular \cite{bohn17}. For instance, a degenerate Fermi gas of polar molecules~\cite{demarco19}, or a dipolar quantum gas with metastable supersolid properties~\cite{tanzi19} were observed recently. The drawback of the static dipole-dipole interactions is that they decay as $1/{d^3}$, that is with $\alpha=3$ the system is not very different from the short-range chain. 

(ii)\textit{Fermionic atoms/molecules with laser-induced dipole moments.}  Already in  2000, Kurizki and collaborators propose to induce dipole-dipole interactions using off-resonant laser excitation. Such interactions are subject to retardation effects, and as such, in addition to $1/{d^3}$ term, they include $1/{d^2}$ and $1/{d^1}$ terms. O'Dell {\it et al.}~\cite{odell00,giovanazzi02,kurizki04} showed that particular configurations of intense off-resonant laser beams can give rise to an attractive $1/d$ interatomic potential.  Such a "gravitational-like" interaction leads to stable Bose-Einstein condensates that are self-bound (without an additional trap) with very interesting properties. Light-induced dipole-dipole interactions were intensively studied in experiments~\cite{low05} and theory~\cite{papadopoulos07}, leading to more recent observations of long-range one-dimensional gravitational-like interactions in a neutral atomic cold gas~\cite{chalony13}, or light induced inverse-square law interactions between nanoparticles~\cite{luis19}. The drawback of this approach is that light induced dipole interactions contain not only a conservative part, but a dissipative part too.  

(iii) \textit{Fermionic atoms/molecules with synthetic Coulomb interactions.} Finally, there is a recent proposal for analogue quantum chemistry simulation~\cite{arguello19}, allowing in particular for synthetic Coulomb interactions between fermionic particles playing role of electrons. This proposal does not belong to the mainstream of quantum computational chemistry~\cite{mcardle20}, which maps fermionic operators to qubits (spins 1/2) via Jordan-Wigner transformations, and uses  quantum devices operating on qubits. Still it has generated a lot of interest and excitement in the community, and first experiments in this direction are on their way~\cite{heinz20}. The challenge here would be to find a regime with attractive interactions.

\section{\label{sec:level7} Methods}

The code that has been used to obtain all the data and Figures for the paper can be found in \cite{code20}.

\section{\label{sec:level8} Conclusions}

In this paper, we studied in detail the bulk topology and the edge states of the long-range Kitaev chain with an AAH potential. In this context, we discussed several algorithms to characterize the bulk topology, and we derived the phase diagram of the system at half filling. We also studied the appearance of MMs and MDMs in the system: in both short-range and long-range systems with an AAH potential, the appearance of MMs and MDMs requires a non-zero critical pairing. The critical values of the pairing depend on the long-range decay exponent $\alpha$, and as such these values are very different for the short and the long-range scenario. We also discovered a peculiar behavior: in the long-range case with AAH potential, MDMs exhibit true crossings at zero energy. We found that these crossings have no effect on the winding number. Nevertheless, we showed that they are adibatically connected to the extended phase with MMs for $\alpha\gg1$, and thus have a direct relation to Majorana physics. Finally, we studied the energy gaps at higher fillings. They are characterized by a 2D Chern invariant, which can efficiently be computed numerically. An interesting behavior appeared in the study of the edge states: the AAH edge modes were dramatically affected by the long-range superconducting pairing, and the bulk-boundary correspondence appeared to be weakened. 

As an outlook, it would be interesting to study the quasi-periodic limit of the system. Quasiperiodic systems are known to exhibit a multifractal energy spectra~\cite{tang86, hiramoto92} and critical, extended or localized wavefunctions depending on the chemical potential amplitude~\cite{wang16}. The multifractal and localization properties of the AAH model have been extensively studied for different generalizations of the model~\cite{hiramoto89, wang16, yao19, satija13, yahyavi19, zeng19}, but the interplay between long-range superconducting pairing and incommensurability has not yet been explored. For the system under study there are still several open questions, such as the effect of the decay exponent $\alpha$ on the Anderson localization-delocalization transitions and an investigation of the multifractal properties of the system's wavefunctions.

\begin{acknowledgments}
We acknowledge support from ERC AdG NOQIA, Spanish Ministry MINECO and State Research Agency AEI (FIDEUA PID2019-106901GB-I00/10.13039 / 501100011033, SEVERO OCHOA No. SEV-2015-0522 and CEX2019-000910-S, FPI), European Social Fund, Fundaci\'o Cellex, Fundaci\'o Mir-Puig, Generalitat de Catalunya (AGAUR Grant No. 2017 SGR 1341, CERCA program, QuantumCAT U16-011424, co-funded by ERDF Operational Program of Catalonia 2014-2020), MINECO-EU QUANTERA MAQS (funded by State Research Agency (AEI) PCI2019-111828-2 / 10.13039/501100011033), EU Horizon 2020 FET-OPEN OPTOLogic (Grant No 899794), and the National Science Centre, Poland-Symfonia Grant No. 2016/20/W/ST4/00314. A.D. acknowledges the Juan de la Cierva program (IJCI-2017-33180) and and the financial support from a fellowship granted by la Caixa Foundation (ID 100010434, fellowship code LCF/BQ/PR20/11770012). T.G. acknowledges financial support from a fellowship granted by ``la Caixa'' Foundation (ID 100010434, fellowship code LCF/BQ/PI19/11690013). U.B. and D.R. acknowledge support by the  ``Cellex-ICFO-MPQ Research Fellows'', a joint program between ICFO and MPQ - Max-Planck-Institute for Quantum Optics, funded by the Fundaci\'o Cellex.
\end{acknowledgments}

\appendix

\section{\label{appendix:construction}Construction of the Bogolibov-de-Gennes momentum representation of the Hamiltonian}

We write the Hamiltonian~\eqref{eq: hamiltonian} in terms of the real-space Bougolibov-de-Gennes basis $\chi = \left(c_{0}, c^\dagger_{0}, c_{1}, c^\dagger_{1}, ..., c_{N-1}, c^\dagger_{N-1}\right)^T$
\begin{equation}
    H = \chi^\dagger H_N \chi
\end{equation}
where:
\begin{equation}
H_N = 
\begin{pmatrix}
A_0 & B & C_2 &\cdots  & -B^\dagger\\
B^\dagger & A_1 & B & \cdots & C_{N-2}\\
\vdots  & \vdots  & \vdots  &  \ddots & \vdots\\
C_{N-2}^\dagger & C_{N-3}^\dagger & C_{N-4}^\dagger & \cdots & B\\
-B & C_{N-2}^\dagger & C_{N-3}^\dagger &  \cdots & A_{N-1},
\end{pmatrix}
\end{equation}
with $A_{j} = -\mu f(j) \sigma_z$, $B = \frac{t}{2}\sigma_z -\Delta i \sigma_y$ and $ C_l = -\frac{\Delta}{d_l^{\alpha}}i\sigma_y$. Here, we have replaced the $l$ of Eq.~\eqref{eq: hamiltonian} by $d_l = \text{min}(l, N-l)$ to take into account the APBC.

From Eq. \eqref{eq:modulation}, we know that the system has a periodicity of $q$ sites, therefore we can rearrange the Hamiltonian into three contributions: $H_{\text{local}}$, which contains the terms within each supercell of $q$ sites, $H_{\text{hop}}$ which contains the hoppings between the adjacent supercells and $H_{\text{l}}$, which connects all the supercells of the system through the long-range superconducting pairing
\begin{eqnarray}\label{eq: A hamiltonian rs}
    H =&& \sum_{u=0}^{L-1} \biggl[ \chi_u^\dagger H_{\text{local}} \chi_u + \left( \chi_u^\dagger H_{\text{hop}} \chi_{u+1} + {\rm h.c.}\right) \nonumber \\
    && + \sum_{l=1}^{L-1}\left( \chi_u^\dagger H_{\text{l}} \chi_{u+l} + {\rm h.c.} \right) \biggl],
\end{eqnarray}
where $\chi_u = \left(c_{qu}, c^\dagger_{qu}, ..., c_{qu+(q-1)}, c^\dagger_{qu+(q-1)}\right)^T$ and $L$ is the number of supercells of the system, namely $L=N/q$. Each contribution to the Hamiltonian is defined as follows
\begin{equation}
H_{\text{local}} = 
\begin{pmatrix}
A_0 & B & C_2 & \cdots & C_{q-1}\\
B^\dagger & A_1 & B & \cdots  & C_{q-2}\\
\vdots  & \vdots  & \vdots  & \ddots & \vdots\\
C_{q-2}^\dagger & C_{q-3}^\dagger & C_{q-4}^\dagger  & \cdots  & B\\
C_{q-1}^\dagger & C_{q-2}^\dagger & C_{q-3}^\dagger & \cdots & A_{q-1}
\end{pmatrix},
\end{equation}
\begin{equation}
H_{\text{hop}} = 
\begin{pmatrix}
0 & 0 & \cdots & 0\\
0 & 0 &  \cdots  & 0\\
\vdots   & \vdots & \ddots & \vdots\\
B' & 0 &  \cdots  & 0
\end{pmatrix},
\end{equation}
and,
\begin{equation}
H_{\text{l}} = 
\begin{pmatrix}
C_{l,0,0} & C_{l,0,1} & \cdots & C_{l,0,q-1}\\
C_{l,1,0} & C_{l,1,1} & \cdots & C_{l,1,q-1}\\
\vdots  & \vdots  & \ddots & \vdots\\
C_{l,q-2,0} & C_{l,q-2,1} & \cdots &  C_{l,q-2,q-1}\\
C_{l,q-1,0} & C_{l,q-1,1} & \cdots & C_{l,q-1,q-1}
\end{pmatrix},
\end{equation}
where $B' = \frac{t}{2}\sigma_z$ and $C_{l,x,y} = -\frac{\Delta}{2d_{l,x,y}^{\alpha}} i\sigma_y$. For the system with APBC, 
\begin{equation}
	d_{l,x,y} = \text{min}\left(lq-(x-y), N-(lq-(x-y))\right).
\end{equation}
Moreover, we explicitly impose APBC by assuming that
\begin{equation}\label{eq: APBC}
	\chi_{u+L} = -\chi_{u}.
\end{equation}
In order to obtain the momentum space Hamiltonian, we need to use the Fourier transformation of the spinor $\chi_u$
\begin{eqnarray}\label{eq: A fourier spinor}
    \chi_u &=& \frac{1}{\sqrt{L}} \sum_{k} e^{iku} \chi_k, \nonumber 
\end{eqnarray}
where $\chi_k= \left(c_{k,0}, c^\dagger_{-k,0}, ..., c_{k, q-1}, c^\dagger_{-k, q-1}\right)^T$. Note that $u \in \{0,...,L-1\}$ denotes the supercell index. 

Combining Eq.~\eqref{eq: A fourier spinor} with Eq.~\eqref{eq: APBC}, we find
\begin{equation}
    \frac{1}{\sqrt{L}} \sum_{k} e^{ik(u+L)} \chi_k = -\frac{1}{\sqrt{L}} \sum_{k} e^{iku} \chi_k,
\end{equation}
which implies that the momentum $k$ is defined as
\begin{equation}
 k = \frac{(2m + 1)\pi}{L},
\end{equation}
where $m \in \{0,1,2,...L-1\}$. \\
The momentum representation of the Hamiltonian takes the form
\begin{eqnarray}
    H &=& \sum_{k} \biggl[ \chi_k^\dagger H_\text{local} \chi_k + \left(e^{ik} \chi_k^\dagger H_\text{hop} \chi_k + {\rm h.c.}\right) + \nonumber \\ &&\sum_{l=1}^{L-1}\left(e^{ikl}\chi_k^\dagger H_{l} \chi_{k} + {\rm h.c.}\right)\biggl],
\end{eqnarray}
where, in order to simplify the expression, we used
\begin{equation}
	\sum_{u=0}^{L-1} e^{i(k-k')u} = L\delta_{kk'}.
\end{equation}

\section{\label{appendix:APBC}Anti-periodic boundary conditions}

If we consider a system with $\beta = p/q$, it has a periodicity of $q$ sites. Then, the total length of the system will be $N=Lq$, where $L$ is the number of supercells. We can take $L$ to be a finite number or consider the system for the limit $L\rightarrow \infty$. Either way, we can still consider different types of boundary conditions. 
Usually, in order to construct the momentum space representation of the Hamiltonian, we need to consider a system with either periodic (PBC) or anti-periodic (APBC) boundary conditions. The long-range pairing terms of the Hamiltonian get canceled if we impose PBC, and that is why we consider APBC when needed in our calculations. It is important to note that, in the infinite limit $L\rightarrow \infty$, the boundary conditions can be neglected. Then, there is no technical difference between considering APBC or PBC.

\section{\label{appendix:block-off}Block-off diagonal infinite Hamiltonian}

We transform the infinite Hamiltonian $H_\text{inf}$ of Eq.~\eqref{eq: hamiltonian inf} as follows
\begin{equation}
H_\text{inf}' = R_y H_\text{inf} R_y^\dagger 
\end{equation}
where $R_y = e^{i\sigma_y \frac{\pi}{2}} \otimes \mathbb{1}_q$. Moreover, we perform a rotation in order to rearrange the basis to the following form
\begin{equation}
	\chi_k' = \left(c_{k,0},  ..., c_{k, q-1}, c^\dagger_{-k,0}, ..., c^\dagger_{-k, q-1}\right)^T.
\end{equation}
We are left with
\begin{equation}
	H = \sum_k \chi_k'^{\dagger}H_{inf}'\chi_k',
\end{equation}
where $H_{inf}'$ is block-off diagonal.

\section{\label{appendix:WBBC} Weak bulk-boundary correspondence}

\begin{figure}[h]
\includegraphics[width=\columnwidth]{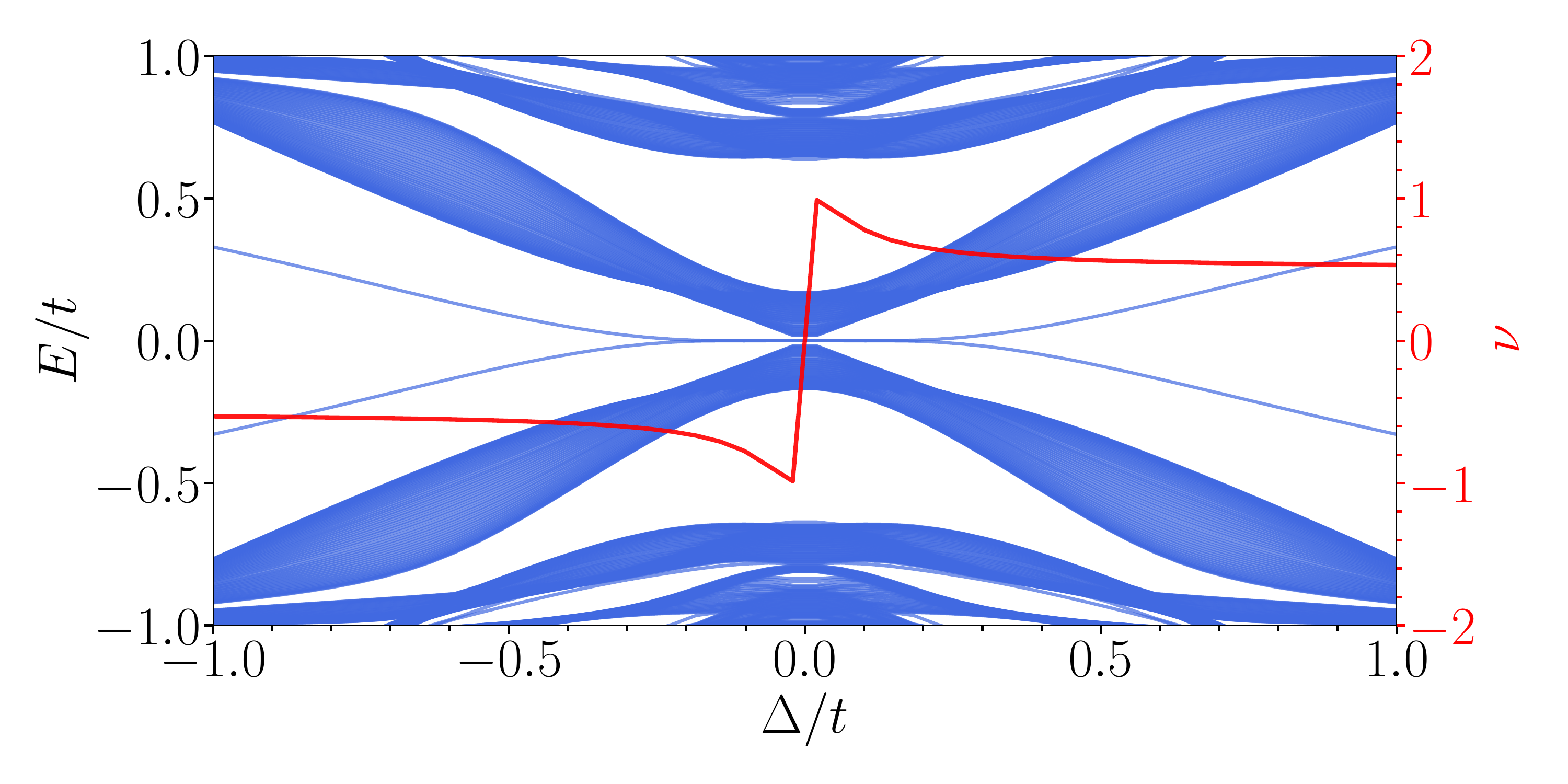}
\caption{\label{fig:weak_bec} Energy spectrum with OBC and real-space winding number for a system with $\mu/t = 0.5$, $\alpha = 0.5$ and $\phi = 0$. We see that the real-space winding number changes sign at $\Delta = 0$, while the edge states exist at both sides of the gap closing. The plot is obtained for a system with size of $q=233$ and $L=1$.}
\end{figure}

Figure~\ref{fig:weak_bec} shows the comparison between the behavior of the edge dynamics and the bulk topology of the long-range system. In particular, by looking at the energy spectrum for OBC we notice that the system exhibits MDMs for any value of $\Delta/t \in [-1,1]$ but $\Delta = 0$, which corresponds to a gap closing, and the region around $\Delta = 0$, for which we find MMs. The existence of MMs for small values of $\Delta$ was already adressed in Figure~\ref{fig: cd_constant}. Here, instead, we are interested in comparing the edge dynamics with the bulk topology. If we look at the real-space winding number for the same system, we see that it takes a positive value $+0.5$ for $\Delta/t > 0$ and a negative value $-0.5$ for $\Delta/t < 0$. This means that at $\Delta = 0$ we have a topological phase transition, but this does not affect the edge dynamics. Therefore, we clearly show how the bulk-boundary correspondence is broken when the sign of the superconducting pairing $\Delta$ is changed.

%\nocite{*}                                                                                                                                                                                                
\bibliographystyle{apsrev4-1}

%\nocite{*}

\end{document}